\documentclass[aps,prl,reprint,superscriptaddress]{revtex4-2}

\usepackage{graphicx}
\usepackage{dcolumn}
\usepackage{bm}

\usepackage[utf8]{inputenc}
\usepackage[T1]{fontenc}
\usepackage{mathptmx}
\usepackage{etoolbox}
\usepackage{soul}
\usepackage{xcolor}
\usepackage{amssymb}
\usepackage{amsmath}
\usepackage{bbm}
\usepackage{hyperref}
\renewcommand{\i}{\mathrm{i}}
\newcommand{\Hsys}{H_\mathrm{sys}}

\newcommand{\Henv}{H_\mathrm{env}}
\newcommand{\Htot}{H_\mathrm{tot}}

\newcommand{\cHsys}{{\cal H}_\mathrm{sys}}
\newcommand{\cHenv}{{\cal H}_\mathrm{env}}

\newcommand{\dd}[1]{\mathrm{d}#1}
\newcommand{\bvec}[1]{\mathbf{#1}}
\newcommand{\ket}[1]{|#1\rangle}

\newcommand{\ptr}[2]{\operatorname{tr}_{#1}\left[ #2 \right]}
\newcommand{\Kadd}[1]{{#1}}

\DeclareSymbolFont{newfont}{OML}{cmm}{m}{it}
\DeclareMathSymbol{\Epsilon}{3}{newfont}{15}
\DeclareMathSymbol{\Varrho}{3}{newfont}{37}

\makeatletter
\def\@email#1#2{%
 \endgroup
 \patchcmd{\titleblock@produce}
  {\frontmatter@RRAPformat}
  {\frontmatter@RRAPformat{\produce@RRAP{*#1\href{mailto:#2}{#2}}}\frontmatter@RRAPformat}
  {}{}
}%
\makeatother
\begin{document}

\title{One-to-one correspondence between Hierarchical Equations of Motion and Pseudomodes for Open Quantum System Dynamics}
\author{Kai Müller}
\affiliation{Institut für Theoretische Physik, Technische Universität Dresden, D-01062 Dresden, Germany.}
\author{Walter T. Strunz}
\affiliation{Institut für Theoretische Physik, Technische Universität Dresden, D-01062 Dresden, Germany.}
\date{\today}

\begin{abstract}
We unite two of the most widely used approaches for strongly damped, non-Markovian open quantum dynamics, the Hierarchical Equations of Motion (HEOM) and the pseudomode method by proving two statements: First, every physical bath correlation function (BCF) that can be written as a sum of $N$ exponential terms 
can be obtained from a \emph{physical} model with $N$ interacting pseudomodes which are damped in Lindblad form. 
Second, for every such BCF there exists a non-unitary, linear transformation which mirrors the evolution of the system-pseudomode state onto the HEOM hierarchy, and vice versa.
Our proofs are constructive and we give explicit expressions for the mirror transformation as well as for the 
pseudomode Lindbladian corresponding to a given exponential BCF. This approach also gives insight and provides elegant derivations of the corresponding Hierarchy of stochastic Pure States (HOPS) method and its nearly-unitary 
version, nuHOPS. Our result opens several avenues for further optimization of non-Markovian open quantum 
system dynamics methods.
\end{abstract}

\maketitle

\paragraph{Introduction}
The dynamics of open quantum systems beyond the Markovian approximation have attracted sustained interest over the past decades \cite{Breuer2007Jan,Review_de_Vega, MengXu}. In many physically relevant settings, the assumption of a memoryless environment breaks down, and the system dynamics are strongly influenced by temporal correlations in the surrounding bath \cite{Weiss2011Nov,Groblacher2015Jul,Dai2025Jan}. \\
A wide variety of numerical and analytical methods has been developed to investigate these non-Markovian open quantum systems \cite{Chin2025Aug, Cerrillo2014Mar, HOPS,nuHOPS,Link_2023,TEMPO,Fowlwer-Wrigth_2022,uniTEMPO,ACE,TEDOPA,TEDOPA_algorithm,Garraway1997Mar,Pleasance2020Oct,Pseudomode_Plenio,pseudomodes_lin,pseudomodes_Feist,ML-MCTDH,carmichael1999statistical,Davydov_review,Davydov_Frank,Tanimura_1990,Tanimura2020}.
A large and practically important subclass of them is based on the fact that the influence of a Gaussian bath is fully characterized by its bath correlation function \cite{Feynman1963Oct}. The original environment can thus be replaced by an effective, extended state space that reproduces the same bath correlation function (BCF) \cite{MengXu}. 
Well-known examples from this subgroup include 
pseudomode methods \cite{Garraway1997Mar,Pleasance2020Oct,pseudomodes_Feist,pseudomodes_Imamoglu,pseudomodes_Yang,pseudomodes_Knorr, Lentrodt20}, chain mappings \cite{Chain_Voijta,Chain_Burghardt, Chain_Plenio,Chain_Feist}, the hierarchical equations of motion (HEOM) \cite{Tanimura2020,Tanimura_1990,Hsieh2018Jan,Jin2008Jun,Batge2021Jun,Debecker2024Oct,Xu2022Nov}, the hierarchy of pure states (HOPS) \cite{HOPS,HOPS_Richard,mesoHOPS} and its improved nearly unitary version nuHOPS \cite{nuHOPS}.
Also some tensor-network-based schemes \cite{TEMPO} like uniTEMPO (uniform time evolving matrix product operator) \cite{uniTEMPO} can be understood in this manner.\\
Here we focus specifically on the relation between the pseudomode approach, and the hierarchical methods (HEOM, HOPS, nuHOPS), and we touch chain mapping approaches later. While they all embed the system into an effective environment, the construction and physical transparency of the environments differ significantly. \\
In the hierarchical methods, the effective environment involves auxiliary degrees of freedom which can be directly constructed from an exponential fit of the BCF.
This direct approach with few parameters allows us to use highly efficient and accurate fit routines \cite{expFittingOverview}.
The price to pay is that -- except in special cases \cite{MengXu} -- a general simple physical intuition for the auxiliary degrees of freedom in hierarchical methods is currently still lacking. Moreover, there is the possibility that the approximate, exponential fits to the physical BCF -- involving complex valued amplitudes, in general -- do not provide positive kernels. In other words, the approximate fits might correspond to spectral densities that are negative at certain frequencies which can lead to instabilities in the evolution \cite{Dunn2019May,Yan2020Nov,Krug2023Dec}.\\
In contrast, the pseudomode approach represents the effective environment as a set of damped harmonic modes -- possibly interacting -- with a clear physical interpretation, which also guarantees a well behaved CPT (completely positive and trace preserving) evolution of the reduced state.
However, fitting a given BCF to an interacting pseudomode model requires advanced methods.
These involve either a direct optimization of all free parameters in the Ansatz Liouvillian \cite{pseudomodes_Feist, pseudomodes_lin}, a HEOM-like fit of the BCF with additional, involved procedures to guarantee a physical model \cite{Pleasance2020Oct,pseudomodes_Yang} or both \cite{Pseudomode_Plenio}. 
As a consequence, it is a long standing open question which bath correlation functions are representable by interacting pseudomodes \cite{Garraway1997Mar,Alford_2_exponentials,Pleasance2020Oct}. Rigorous results have hitherto been restricted to the special case of two pseudomodes only \cite{Alford_2_exponentials,Pleasance2020Oct}.\\

\begin{figure}
    \centering
    \includegraphics[width=0.9\linewidth]{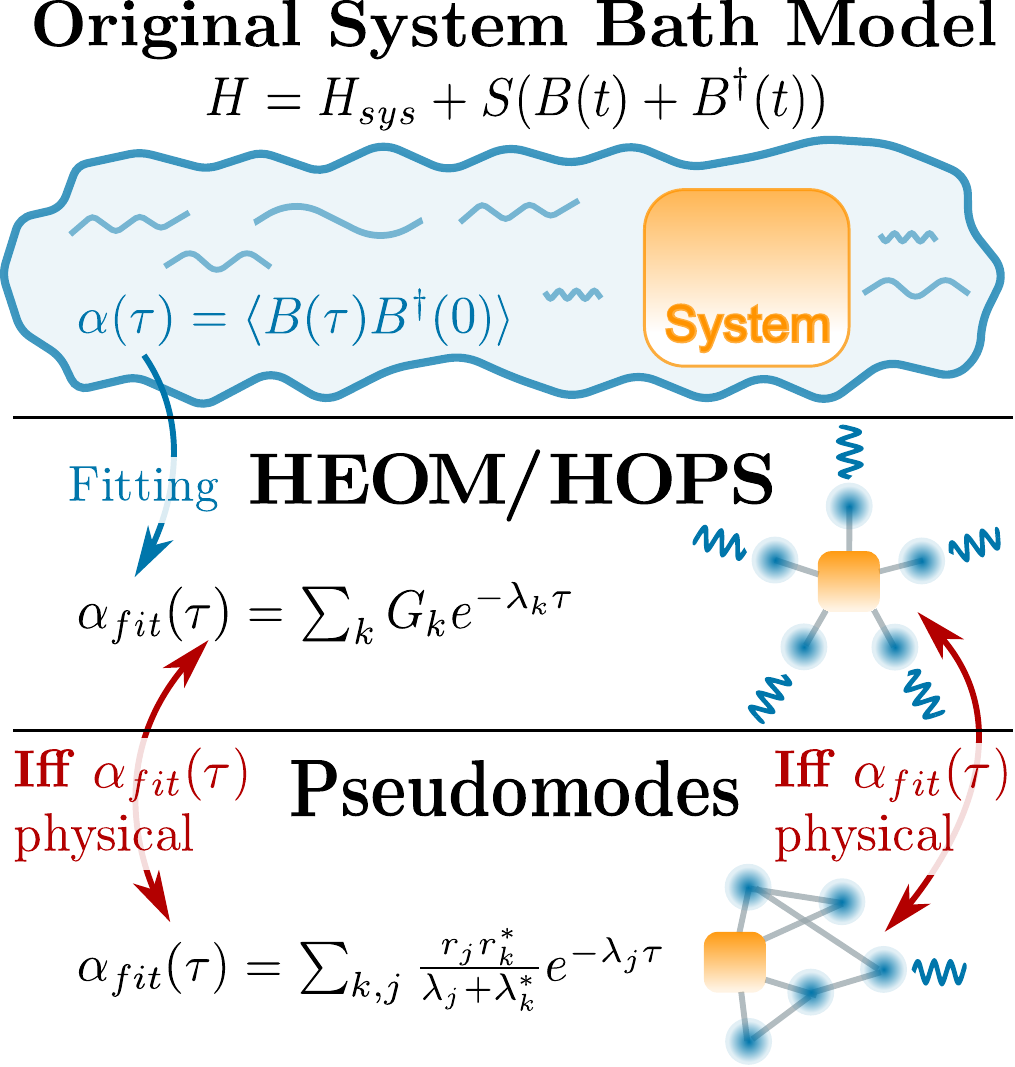}
    \caption{Schematic overview of our results. Through an exponential fit of the BCF $\alpha(\tau)$ the hierarchical methods HEOM and HOPS replace an arbitrary Gaussian bath by a set of auxiliary, bosonic modes in a star geometry. We show that iff this fit corresponds to a physical BCF (that is, a non-negative corresponding spectral density) one can find an equivalent representation of the environment in terms of interacting pseudomodes in Lindblad form. Surprisingly, a model where only one the modes is damped is sufficiently general to capture all possible BCFs.}
    \label{fig:placeholder}
\end{figure}

In this Letter we give a conclusive answer to this question: we prove that \emph{every} physical BCF that can be expressed as a sum of $N$ complex exponential terms can be represented by a pseudomode model with $N$ interacting pseudomodes, generalizing all previous results. 
Our proof is constructive and allows to directly obtain the desired pseudomode Hamiltonian and Lindblad operators from a given (physical) BCF in sum-of-exponentials form.
Additionally, our newly found insights suggest an Ansatz for the exponential fit that parametrizes only the (full) space of physical exponential BCFs, while requiring the same number of parameters as a direct fit.
Such a fit allows us to obtain pseudomode models easily, and on top, guarantees the stability of HEOM calculations by combining the best of both worlds.\\
While these results already establish the equivalence of HEOM and the interacting pseudomodes method on the level of the reduced system, our studies reveal even more:
in the second part of this Letter we show that 
whenever the fitted BCF is physical, there always exists a linear transformation that maps the state of system plus pseudomodes to the corresponding hierarchy of HEOM states and vice versa. 
A transformation of this kind was recently found by \Kadd{M.} Xu \textit{et.al.} in Ref.~\cite{MengXu} for BCFs with real, positive amplitudes of the exponential contributions.
Here we provide a transparent derivation of this one-to-one correspondence between pseudomodes and hierarchical methods (HEOM, HOPS) in the most general case. For all physical BCFs explicit expressions of the transformation can be given, which allows us to offer a physically appealing interpretation of that transformation: analogous to the physics of a beam splitter, the hierarchy emerges as a peculiar mirror image of the pseudomodes.

\paragraph{Gaussian environment model -- }
The problem we aim to solve is to find the reduced evolution of $\rho_{\mathrm{sys}}(t) = \ptr{\mathrm{env}}{\rho_{\mathrm{tot}}(t)}$, where $\rho_{\mathrm{tot}}(t)$ evolves under a fundamental system-environment Hamiltonian. In interaction picture with respect to the free environmental Hamiltonian $H_{\mathrm{env}}^{\mathrm{(orig)}}$ the full system-environment dynamics is determined through
\begin{equation}\label{eq:FundamentalGaussmodel}
    \Htot(t) = \Hsys + S\otimes (B(t) +B^\dagger(t)),
\end{equation}
with $B(t) = \exp\{\i H_{\mathrm{env}}^{(\mathrm{orig})} t\} B \exp\{-\i H_{\mathrm{env}}^{(\mathrm{orig})} t\}$ and arbitrary system operators $\Hsys$, $S$ .
As usual, we assume that $B(t)$ describes a stationary Gaussian bath. This is uniquely characterized by its BCF or, equivalently, by its spectral density $J(\omega)$ through
\begin{equation}\label{eq:BCFexact}
    \alpha(t-s) = \langle B(t) B^\dagger(s) \rangle= \frac{1}{2\pi}\int_{-\infty}^\infty J(\omega) e^{-i\omega(t-s)} \dd\omega.
\end{equation}

Note here that $J(\omega)$ is an effective spectral density, 
permitting negative-frequency oscillators as required for finite temperature baths.

Being a correlation function implies that $\alpha(\tau)$ is a \emph{positive semi-definite kernel}. Therefore,
$J(\omega)$ must 
be strictly non-negative for all frequencies,
\begin{equation}\label{eq:positiveKernel}
    J(\omega) =  \int_{-\infty}^\infty \alpha(\tau)e^{i\omega\tau}\dd{\tau} \geq 0\quad\forall\omega\in\mathbb{R}.
\end{equation}
We refer to BCFs that satisfy Eq.~\eqref{eq:positiveKernel} as {\emph{physical}} BCFs.

\paragraph{Hierarchies and pseudomodes -- }
Hierarchical methods like HEOM, HOPS or nuHOPS \cite{HOPS, HOPS_Richard, nuHOPS} are based on the replacement of the exact BCF of the environment \eqref{eq:BCFexact} by a fitted sum of exponential functions $\alpha(\tau)\approx\alpha_{\mathrm{exp}}(\tau)$, with
\begin{equation}\label{eq:exponentialcorrelation}
    \alpha_{\mathrm{exp}}(\tau) = \sum_{j=1}^{N} G_j \exp\{-\lambda_j \tau\},\qquad\;(\tau\ge 0)
\end{equation}
and $\alpha_{\mathrm{exp}}(-\tau) = \alpha_{\mathrm{exp}}(\tau)^*$. In this Letter, with a slight abuse of terminology, we refer to BCFs of this form as {\emph{exponential}} BCFs. We stress that the amplitudes $G_j$ need not be real and positive to
represent a positive kernel. A prime example of this kind is the $N=2$-band-gap model \cite{Garraway1997Mar}, where $G_1$ is real positive, but $G_2<0$, such that $J(\omega)\ge 0 $ has a real zero. Indeed, in general, the fit parameters are allowed to be complex valued, $G_j,\lambda_j\in \mathbb{C}$, with $\operatorname{Re}({\lambda_j})>0$. Compared to an Ansatz with real, positive $G_j\in\mathbb{R}_{\geq0}$, using complex amplitudes results in significantly fewer exponential terms for a given accuracy \cite{pseudomodes_lin,quasi_lindblad_abanin_fermions}. Due to sophisticated fitting methods \cite{expFittingOverview}, \Kadd{they can be obtained} with a remarkable efficiency \cite{Yu2026Jan}.
Based on Eq.~\eqref{eq:exponentialcorrelation}, the HEOM approach provides a hierarchy of coupled equations for auxiliary density matrices $\rho^{(\bvec{n},\bvec{m})}$, whose root state 
${\rho}^{(\bvec{0},\bvec{0})}(t)=\rho_{\mathrm{sys}}(t)$ is the desired quantum state of the open system \cite{Tanimura2020}.
It was noted in \cite{dissiponsAlex,dissiponsAndrew} that the coupled hierarchical equations take a concise form by identifying $\rho^{(\bvec{n},\bvec{m})} = \langle\bvec{n}\mathrm{|}\Varrho\ket{\bvec{m}}$ with number states $\ket{\bvec{n}}=\ket{n_1,\ldots,n_N}$ of $N$ auxiliary bosonic modes. 
Then, the HEOM equations take the appealing form
\begin{equation}\label{eq:HEOM}
    \begin{split}
        \partial_t\Varrho =& -\i \left[\Hsys,\Varrho\right]  - \sum_{j=1}^{N} \left(\lambda_jb_j^\dagger b_j\Varrho + \lambda_j^*\Varrho b_j^\dagger b_j\right)\\
        &-i \sum_{j=1}^{N} \left(\sqrt{G_j}\left[S, b_j\Varrho\right] + \sqrt{G_j}^*\left[S,\Varrho b_j^\dagger\right]\right)\\& -\i \sum_{j=1}^{N}\left(\sqrt{G_j}Sb_j^\dagger\Varrho  - \sqrt{G_j}^* \Varrho b_jS\right).
    \end{split}
\end{equation}
In the following, we will refer to these auxiliary bosons
as {\emph{dissipons}}, and their connection to the physical bosonic environment will be established soon. Thus, we equip the dissipon-Hilbert space with usual annihilation and creation operators $b_i,b^\dagger_j$, such that $[b_i, b^\dagger_j]=\delta_{ij}$ and HEOM \eqref{eq:HEOM} turns into an evolution equation for an operator $\Varrho(t)$ in the joint system-dissipon Hilbert space. \Kadd{Note that the dissipons only interact with the system and not amongst themselves, leading to the star geometry shown in Fig.~\ref{fig:placeholder}.} We repeat that the usual hierarchical form of HEOM arises from an expansion of \eqref{eq:HEOM} in dissipon-Fock states: $\rho^{(\bvec{n},\bvec{m})} = \langle\bvec{n}\mathrm{|}\Varrho\ket{\bvec{m}}$.\\

By contrast, the pseudomode method is based on a set of damped bosonic modes $a_k$, that may be linearly coupled among each other. The original bosonic environment from \eqref{eq:FundamentalGaussmodel} is replaced by these pseudomodes such that the total evolution equation for a density operator in the system-pseudomode Hilbert space can be given in Lindblad form
\begin{equation}\label{eq:defPseudomode}
    \begin{split}
    \dot{\rho} =& -i[H_{\mathrm{pm}},\rho] + \sum_k L_k\rho L_k^\dagger -\frac{1}{2}\{L_k^\dagger L_k, \rho\},\\
        H_{\mathrm{pm}} =& H_{\mathrm{sys}} + S\otimes\sum_k(g^*_ka_k+g_ka_k^\dagger)  + \sum_{k,k'}h_{k,k'}a_k^\dagger a_{k'}.
    \end{split}
\end{equation}
Here, $L_k = \sum_j \Gamma_{k,k'} a_{k'}$ and $h$ is a Hermitian matrix, containing the pseudomode frequencies and allowing for couplings among them.
For our proofs it turns out to be more convenient to replace the reduced description in terms of a Lindblad master equation by an operator quantum stochastic Langevin equation. This relation is well established \cite{Parthasarathy1992, Gardiner2004}. Thus, we work with a total Hilbert space containing the system ($\cHsys$), and the full environment $\cHenv$, which contains the pseudomodes plus their respective Markovian bath degrees of freedom.

As before, in this total Hilbert space $\cHsys\otimes\cHenv$ we transform to an interaction picture with respect to the full environment, to write the total Hamiltonian in the form of Eq.~\eqref{eq:FundamentalGaussmodel}, as shown in the Supplemental Material (SM) \cite{supplement}.
With $A(t) = \sum_kg_k^* a_k(t)$
we find
\begin{equation}\label{eq:pseudomodeOU}
    \begin{split}
        H_{\mathrm{pm,tot}} =& H_{\mathrm{sys}} + S\otimes(A(t)+A^\dagger(t)),\\
        \dot{a}_k(t) =& \sum_{k'}-(ih_{k,k'} +\frac{1}{2}(\Gamma^\dagger\Gamma)_{k,k'})a_{k'} + \Gamma^\dagger_{k,k'}{\xi}_{k'}(t).
    \end{split}
\end{equation}
The operators $\xi_{k}(t)$ are composed of the annihilation operators of the Markovian baths \cite{supplement} and describe operator white noise, with vacuum correlation function \cite{Parthasarathy1992,Gardiner2004} $\langle \xi_k(t)\xi_{k'}^\dagger(s)\rangle_{\mathrm{vac}} = \delta_{k,k'}\delta(t-s)$.
From Eq.~\eqref{eq:pseudomodeOU}, the Heisenberg picture pseudomodes $a_k(t)$ are components of a multivariate operator-valued Ornstein-Uhlenbeck (OU) process and the BCF of the pseudomode environment can be obtained explicitly as 
$\alpha_{\mathrm{pm}}(t-s)=\langle A(t)A^\dagger(s)\rangle=\bvec{g}^\dagger\exp\left(-(\Gamma^\dagger\Gamma/2+\i h)(t-s)\right)
\bvec{g}$ \cite{supplement, pseudomodes_lin},
with a vector $\bvec{g} = (g_1,g_2,\ldots)^T$. 
In order to obtain the reduced state evolution of the original model \eqref{eq:FundamentalGaussmodel} from the pseudomode approach \eqref{eq:defPseudomode}, one needs to find matrices $h, \Gamma$, and coupling constants $\bvec{g}$ that approximately reproduce the BCF \eqref{eq:BCFexact}.

\paragraph{Proof of pseudomode representability --}
In the following we show that if the BCF is given (or approximated) in the form of Eq.~\eqref{eq:exponentialcorrelation} and satisfies the positive kernel condition \eqref{eq:positiveKernel}, it is always possible to find $h,\,\Gamma, \bvec{g}$ such that $\alpha_{\mathrm{pm}}(\tau) = \alpha_{\mathrm{exp}}(\tau)$.
The full details of the proof are given in the SM \cite{supplement} -- here we provide a summary of the most important findings and a discussion of the implications.\\

As discussed, physical models of type \eqref{eq:exponentialcorrelation} require a non-negative spectral density. We show in the SM \cite{supplement} that this property alone guarantees that the amplitudes $G_j$ in Eq.~\eqref{eq:exponentialcorrelation} can be written in the form
\begin{align}\label{eq:G_kCondition}
    G_j = \sum_{k=1}^{N}\frac{r_jr_k^*}{\lambda_j+\lambda_k^*},
\end{align}
where $\bvec{r} = (r_1,\ldots,r_N)^T$ is a complex vector.
This form of $G_j$, in turn, as explained in the SM, is sufficient to guarantee the existence of a corresponding \Kadd{pseudomode model \eqref{eq:pseudomodeOU}.\\ 
In the following, we show how its Lindbladian can be constructed from the knowledge of the parameters in Eq.~\eqref{eq:G_kCondition} and refer to \cite{supplement} for the proof.} 
\Kadd{The first step is }a diagonalization of the positive matrix $P_{jk} = r_j r_k^*/\left(\sqrt{G_jG_k^*}(\lambda_j + \lambda_k^*)\right)$, such that
$P=W\mathcal{D}W^\dagger$ with diagonal $\mathcal{D}$ and unitary $W$. 
\Kadd{This allows us to define}
\begin{equation}
    V = U\mathcal{D}^{-1/2}W^\dagger,
\end{equation}
such that $(V^ \dagger V)^ {-1} =P$. Here, $U$ is an \emph{arbitrary} unitary matrix that changes the 
Lindbladian but not the resulting correlation function.
This unitary freedom can be exploited to bring the Lindbladian into a form that is numerically appealing.
For example, we find that it is surprisingly always possible to choose $U$ in such a way that \emph{only one} of the $N$ pseudomodes is damped, allowing for particularly efficient trajectory unravelings.
In this case we find the following expressions for the Hamiltonian and the Lindblad operator(s)
\begin{equation}\label{eq:parametersLetter}
    \begin{split}
        h =& \frac{1}{2\i}\left(V\lambda V^{-1}- \mathrm{h.c.}\right),\\
        \Gamma^\dagger =& \kappa \bvec{e}_1\bvec{e}_1^\dagger,\\
         g_k =& \sum_jV_{kj}^{-1}\sqrt{G_j^*}
    \end{split}
\end{equation}
where explicit expressions for $U,\,\kappa \in\mathbb{R}_{\geq 0}$ are given in the SM \cite{supplement}.
Although for this choice of $U$ only the first mode in Eq.~\eqref{eq:pseudomodeOU} is damped, we can prove \cite{supplement} that it is nevertheless sufficiently general to represent all physical BCFs of the form \eqref{eq:exponentialcorrelation}.\\
Let us discuss some additional implications of our proof. As mentioned before, it is constructive and allows to determine $h,\,\Gamma, \bvec{g}$ for a given physical BCF in exponential form \eqref{eq:exponentialcorrelation}. The only steps in this process that are not made explicit in our proof \cite{supplement} are the spectral factorization of $J(\omega)=|\nu(\omega)|^2$ and the partial fraction decomposition of $\nu(\omega)$.
However, efficient methods exist for both of these tasks \cite{Sayed2001Sep,Mahoney1983Sep}.
Naturally, the spectral factorization is only applicable to non-negative spectral densities, yet a direct fit with the exponential Ansatz \eqref{eq:exponentialcorrelation} does not guarantee such a positive spectral density.
From a pragmatic point of view, for unphysical bath correlation fits the corresponding dynamics can still be computed using HEOM or a non-hermitian/quasi-Lindblad generalization of the pseudomode method \cite{quasi_lindblad_Lin,quasi_lindblad_abanin_fermions,Pleasance2020Oct}, but one risks instabilities due the loss of the CPT property \cite{Dunn2019May,Yan2020Nov,Krug2023Dec} \Kadd{and precludes the use of quantum trajectory methods \cite{Wiseman2010,Dalibard1992Feb,Carmichael1993}}. \\
Our derivation suggests an alternative to this direct fit. Using Eq.~\eqref{eq:G_kCondition}, which holds if and only if $J(\omega)$ is non-negative, one may parametrize the BCF right from the start as
\begin{equation}\label{eq:posFitAnsatz}
    \alpha_{\mathrm{pos}}(\tau) = \sum_j \left(\sum_k\frac{r_jr_k^*}{\lambda_j+\lambda_k^*}\right)e^{-\lambda_j\tau},
\end{equation}
and optimize the values of $r_k,\lambda_k \in\mathbb{C}$, $\operatorname{Re}(\lambda_k)\geq0$. This optimization problem has the same number of free parameters as the direct fit according to Eq.~\eqref{eq:exponentialcorrelation}, but only samples physical BCFs.
Given the parameters $r_i,\,\lambda_i$ one can directly construct a pseudomode model \cite{supplement}, thus obtaining a guaranteed CPT evolution.\\
Additionally, it is an interesting observation that a pseudomode Ansatz with only one damped mode (and additional coupled, but undamped modes) is already sufficiently general to capture all representable BCFs. 
The presence of only a single damped mode is reminiscent of the chain mapping \cite{Chain_Burghardt,Chain_Feist}, where the environment is mapped onto a chain of modes, with only the last one being coupled to a Markovian bath.
We find, however, that in general couplings between all modes are required and we prove in the SM \cite{supplement} that there are physical BCFs of the form \eqref{eq:exponentialcorrelation} which can not be represented by $N$ modes in a chain like topology with one damped mode. \\
\Kadd{Finally, the parameters in Eq.\eqref{eq:parametersLetter} are not unique, in the sense that there are generally multiple pseudomode models that lead to the same BCF. 
We discuss the freedoms involved in the SM, Sec.~III D \cite{supplement}. Making these freedoms explicit opens the door for exploiting them for numerical optimization.}

\paragraph{Non-unitary dissipon transformation and its beam splitter motivation --}
In the following, we show that the connection between the two approaches \Kadd{is even deeper.}
We are able to find a linear (invertible) \emph{dissipon transformation} $D(t)$ that mirrors the total pseudomode state onto the corresponding dissipon modes of HEOM and HOPS.\\
The meaning of this transformation can be made transparent with a close analogy.
In quantum optics, a beam splitter is a device that splits an incoming light mode (operators $a,a^\dagger$) into two outgoing modes: a transmitted 
(transmission $T$) 
and a reflected mode 
(reflection $R=1-T$). 
This is achieved unitarily by also taking into account the unoccupied (vacuum) second incoming mode (operators $b,b^\dagger$) of the beam splitter. The corresponding two-mode unitary map can be chosen to be 
\begin{equation}\label{eq:beam_splitter_U}
    U_{\mathrm{BS}}= \exp\{v (b^\dagger a - b a^\dagger)\},
\end{equation}
with the angle $v$ determining the mixing $\sqrt{T}=\cos v$ and $\sqrt{R}=\sin v$. 
The beam splitter entangles the incoming mode $a$ with the vacuum $b$ mode,
and whatever happens to the incoming $a$-mode is mirrored onto the outgoing modes with the help of the second (vacuum) $b$-mode.
These considerations give some intuition for the dissipon transformation we discuss in the following.
First, as for the beam splitter, we need to enlarge the pseudomode Hilbert space of operators $a_k(t)$ by an additional set of $N$ harmonic dissipon modes with annihilation (creation) operators $b_j$ ($b_j^\dagger$),  such that the total Hilbert space is now given by $\cHsys\otimes\cHenv\otimes{\cal{H}}_\mathrm{diss}$. These dissipon modes are unoccupied (vacuum state $|\mathrm{vac}\rangle_\mathrm{diss}$) in the beginning. 
The system-pseudomode state $|\Psi(t)\rangle$ satisfies a standard Schr\"odinger equation with total pseudomode Hamiltonian \eqref{eq:pseudomodeOU}. Tracing over the Markovian baths leads to the Lindblad master equation \eqref{eq:defPseudomode}. We now extend this state by
the dissipon vacuum, $|\Psi(t)\rangle|\mathrm{vac}\rangle_\mathrm{diss}$ and mirror the time dependent, decaying pseudomodes onto the dissipon modes through
the non-unitary "beam splitter" dissipon transformation
\begin{equation}\label{eq:defDissipon}
    D(t) = 
\exp\left(\sum_{jk}b_j^\dagger V_{jk}^{-1}a_k(t)\right).
\end{equation}
Here, the matrix $V$ is the one from the Ornstein-Uhlenbeck transformation above, and the pseudommodes ${a}_k(t)$ evolve according to Eq.~\eqref{eq:pseudomodeOU}). 
We thus define a system-pseudomode-dissipon state
\begin{equation}\label{eq:dissipon_transformation}
    \ket{\Phi_t} = D(t)\ket{\Psi(t)}|\mathrm{vac}\rangle_\mathrm{diss}.
\end{equation}
Obviously, $D$ is closely related to the beam splitter unitary (\ref{eq:beam_splitter_U}), yet being non-unitary, there are crucial differences, leading to some unintuitive properties of the transformation. 
Importantly, with $_\mathrm{diss}\langle\mathrm{vac}|b_j^\dagger = 0$ we see that $_\mathrm{diss}\langle\mathrm{vac}| D =\; _\mathrm{diss}\!\langle\mathrm{vac}| \mathbbm{1} = \;_\mathrm{diss}\!\langle\mathrm{vac}|$ and the original pseudomode state can be reconstructed from the entangled state by projection onto the dissipon vacuum $\ket{\Psi(t)} = \;_\mathrm{diss}\!\langle\mathrm{vac|}\Phi(t)\rangle$.

We show in the Supplemental Material \cite{supplement} that the dynamics of the transformed state follows a non-unitary dynamics in the extended Hilbert space of system, environment, and dissipon given by the Schr\"odinger-type equation
\begin{equation}\label{eq:dissipon_pseudomode}
    \begin{split}
        \partial_t |\Phi_t\rangle
   =& \Bigg[  -\i \left(\Hsys + S\otimes A^\dagger(t) + S\sum_{j=1}^ N\sqrt{G_j}\left(b_j+b_j^\dagger\right)\right)\\
   &\quad- \sum_{j=1}^N\lambda_j  b_j^\dagger b_j \Bigg]\, |\Phi_t\rangle.
    \end{split}
\end{equation}
In Eq.~\eqref{eq:dissipon_pseudomode} the original pseudomode environment is still present through the term $S\otimes A^\dagger(t)$. 
However, we are ultimately interested in the reduced system state, obtained by tracing over $\cHenv$. 
After the dissipon transformation, 
tracing over the (pseudomode) environment we obtain the reduced system-dissipon state $\rho_{SD}(t) = \ptr{\mathrm{env}}{|\Phi_t\rangle\!\langle\Phi_t|}$,
from which we obtain the true system state by projection onto the dissipon vacuum,
\begin{equation}\label{eq:projected_state}
    \rho_{sys}(t) =   \;_\mathrm{diss}\!\langle\mathrm{vac}|\rho_{SD}(t) |\mathrm{vac}\rangle_\mathrm{diss}.
\end{equation}
Now, depending on how we take the partial trace $\ptr{\mathrm{env}}{\cdot}$, the different hierarchical methods emerge naturally from Eq.~\eqref{eq:dissipon_pseudomode}. The details are again shown in the SM \cite{supplement}.\\
To obtain HEOM we utilize that under the trace $\ptr{\mathrm{env}}{\cdot}$, the $A(t)^\dagger$ ($A(t)$) operator acting from the "left" ("right") can be replaced by the dissipon creation operators $\sum_j\Tilde{G}_j^{(1)*}b_j^\dagger$ ($\sum_j\Tilde{G}_j^{(1)}b_j$) acting from the "right" ("left"). Then all reference to the original environmental degrees of freedom disappear and we arrive at Eq.~\eqref{eq:HEOM}. 
The dissipons $b_k$ have taken over from the pseudomodes $a_k(t)$ and the reduced state is obtained by projecting onto the dissipon vacuum as in (\ref{eq:projected_state}).\\
If we instead explicitly take the trace in a basis of (unnormalized) coherent states $\|\bvec z\rangle$, the operator $A^\dagger(t)$ is replaced by a scalar noise due to $-i\langle \bvec{z}\|A^\dagger(t) = z_t^*\langle \bvec{z}\|$ and we obtain the (linear) HOPS equations as a stochastic unraveling of Eq.~\eqref{eq:dissipon_pseudomode} \cite{supplement},
in a form observed in \cite{dissiponsAlex}.
The noise $z_t^*$ is given as a sum of (scalar) OU-processes with the autocorrelation function $\langle z_tz_s^*\rangle = \alpha_{\mathrm{exp}}(t-s)$. The non-linear version of HOPS and nuHOPS also follow easily from this approach, as shown in the SM. 

\paragraph{Conclusions} In summary, we have established a one-to-one correspondence between the hierarchical and the pseudomode methods via the dissipon transformation Eqs.~\eqref{eq:defDissipon},\eqref{eq:dissipon_transformation}.
Given this one-to-one correspondence 
, the crucial question at this stage is under which conditions one approach is (numerically) more advantageous than the other. 
Some numerical results compared for the special case of real, positive $G_j$ hint towards preferring HEOM for weakly non-Markovian regimes, yet this seems to change for very strongly non-Markovian regimes \cite{Sukharnikov2026Feb} -- here, further investigations are called for.
For the pseudomode approach, our derivations make the freedoms involved in selecting a Lindbladian for a given BCF explicit \cite{supplement}. An important direction for future research is to exploit these freedoms to optimize the Lindbladian in terms of its numerical performance.\\
Furthermore, as a useful by-product our proof suggests the use of the Ansatz \eqref{eq:posFitAnsatz} for a fit of the BCF in terms of exponentials, which has the same number of parameters as the naive exponential Ansatz \eqref{eq:exponentialcorrelation}. Yet, it parametrizes the entire space of {\emph{physical}, exponential BCFs, it has a direct pseudomode representation and guarantees the CPT dynamics of the reduced state. Integrating this Ansatz into modern algorithms for exponential fitting could be another fruitful direction for further studies.\\
Finally, our results raise the question whether other approaches that rely on the construction of effective environments, like uniTEMPO \cite{uniTEMPO}, fall within the same equivalence class  or can be proven to go beyond it.

\section*{Acknowledgement}
We thank Meng Xu, Valentin Link and Alexander Eisfeld for fruitful discussions.

\bibliography{literature}

\begin{thebibliography}{69}%
\makeatletter
\providecommand \@ifxundefined [1]{%
 \@ifx{#1\undefined}
}%
\providecommand \@ifnum [1]{%
 \ifnum #1\expandafter \@firstoftwo
 \else \expandafter \@secondoftwo
 \fi
}%
\providecommand \@ifx [1]{%
 \ifx #1\expandafter \@firstoftwo
 \else \expandafter \@secondoftwo
 \fi
}%
\providecommand \natexlab [1]{#1}%
\providecommand \enquote  [1]{``#1''}%
\providecommand \bibnamefont  [1]{#1}%
\providecommand \bibfnamefont [1]{#1}%
\providecommand \citenamefont [1]{#1}%
\providecommand \href@noop [0]{\@secondoftwo}%
\providecommand \href [0]{\begingroup \@sanitize@url \@href}%
\providecommand \@href[1]{\@@startlink{#1}\@@href}%
\providecommand \@@href[1]{\endgroup#1\@@endlink}%
\providecommand \@sanitize@url [0]{\catcode `\\12\catcode `\$12\catcode
  `\&12\catcode `\#12\catcode `\^12\catcode `\_12\catcode `\%12\relax}%
\providecommand \@@startlink[1]{}%
\providecommand \@@endlink[0]{}%
\providecommand \url  [0]{\begingroup\@sanitize@url \@url }%
\providecommand \@url [1]{\endgroup\@href {#1}{\urlprefix }}%
\providecommand \urlprefix  [0]{URL }%
\providecommand \Eprint [0]{\href }%
\providecommand \doibase [0]{https://doi.org/}%
\providecommand \selectlanguage [0]{\@gobble}%
\providecommand \bibinfo  [0]{\@secondoftwo}%
\providecommand \bibfield  [0]{\@secondoftwo}%
\providecommand \translation [1]{[#1]}%
\providecommand \BibitemOpen [0]{}%
\providecommand \bibitemStop [0]{}%
\providecommand \bibitemNoStop [0]{.\EOS\space}%
\providecommand \EOS [0]{\spacefactor3000\relax}%
\providecommand \BibitemShut  [1]{\csname bibitem#1\endcsname}%
\let\auto@bib@innerbib\@empty
\bibitem [{\citenamefont {Breuer}\ \emph {et~al.}(2007)\citenamefont {Breuer},
  \citenamefont {Petruccione}, \citenamefont {Breuer},\ and\ \citenamefont
  {Petruccione}}]{Breuer2007Jan}%
  \BibitemOpen
  \bibfield  {author} {\bibinfo {author} {\bibfnamefont {H.-P.}\ \bibnamefont
  {Breuer}}, \bibinfo {author} {\bibfnamefont {F.}~\bibnamefont {Petruccione}},
  \bibinfo {author} {\bibfnamefont {H.-P.}\ \bibnamefont {Breuer}},\ and\
  \bibinfo {author} {\bibfnamefont {F.}~\bibnamefont {Petruccione}},\
  }\href@noop {} {\emph {\bibinfo {title} {The Theory of Open Quantum
  Systems}}}\ (\bibinfo  {publisher} {Oxford University Press},\ \bibinfo
  {address} {Oxford, England, UK},\ \bibinfo {year} {2007})\BibitemShut
  {NoStop}%
\bibitem [{\citenamefont {de~Vega}\ and\ \citenamefont
  {Alonso}(2017)}]{Review_de_Vega}%
  \BibitemOpen
  \bibfield  {author} {\bibinfo {author} {\bibfnamefont {I.}~\bibnamefont
  {de~Vega}}\ and\ \bibinfo {author} {\bibfnamefont {D.}~\bibnamefont
  {Alonso}},\ }\bibfield  {title} {\bibinfo {title} {Dynamics of
  non-{M}arkovian open quantum systems},\ }\href
  {https://doi.org/10.1103/RevModPhys.89.015001} {\bibfield  {journal}
  {\bibinfo  {journal} {Rev. Mod. Phys.}\ }\textbf {\bibinfo {volume} {89}},\
  \bibinfo {pages} {015001} (\bibinfo {year} {2017})}\BibitemShut {NoStop}%
\bibitem [{\citenamefont {Xu}\ \emph {et~al.}(2026)\citenamefont {Xu},
  \citenamefont {Vadimov}, \citenamefont {Stockburger},\ and\ \citenamefont
  {Ankerhold}}]{MengXu}%
  \BibitemOpen
  \bibfield  {author} {\bibinfo {author} {\bibfnamefont {M.}~\bibnamefont
  {Xu}}, \bibinfo {author} {\bibfnamefont {V.}~\bibnamefont {Vadimov}},
  \bibinfo {author} {\bibfnamefont {J.~T.}\ \bibnamefont {Stockburger}},\ and\
  \bibinfo {author} {\bibfnamefont {J.}~\bibnamefont {Ankerhold}},\ }\bibfield
  {title} {\bibinfo {title} {Colloquium: Simulating non-{Markovian} dynamics in
  open quantum systems},\ }\bibfield  {journal} {\bibinfo  {journal} {Rev. Mod.
  Phys.}\ }\href {https://doi.org/10.1103/w3nw-hbjc} {10.1103/w3nw-hbjc}
  (\bibinfo {year} {2026})\BibitemShut {NoStop}%
\bibitem [{\citenamefont {Weiss}(2011)}]{Weiss2011Nov}%
  \BibitemOpen
  \bibfield  {author} {\bibinfo {author} {\bibfnamefont {U.}~\bibnamefont
  {Weiss}},\ }\href {https://doi.org/10.1142/8334} {\emph {\bibinfo {title}
  {Quantum Dissipative Systems}}},\ Series In Modern Condensed Matter Physics\
  (\bibinfo  {publisher} {World Scientific Publishing Company},\ \bibinfo
  {address} {Singapore},\ \bibinfo {year} {2011})\BibitemShut {NoStop}%
\bibitem [{\citenamefont {Gr{\ifmmode\ddot{o}\else\"{o}\fi}blacher}\ \emph
  {et~al.}(2015)\citenamefont {Gr{\ifmmode\ddot{o}\else\"{o}\fi}blacher},
  \citenamefont {Trubarov}, \citenamefont {Prigge}, \citenamefont {Cole},
  \citenamefont {Aspelmeyer},\ and\ \citenamefont
  {Eisert}}]{Groblacher2015Jul}%
  \BibitemOpen
  \bibfield  {author} {\bibinfo {author} {\bibfnamefont {S.}~\bibnamefont
  {Gr{\ifmmode\ddot{o}\else\"{o}\fi}blacher}}, \bibinfo {author} {\bibfnamefont
  {A.}~\bibnamefont {Trubarov}}, \bibinfo {author} {\bibfnamefont
  {N.}~\bibnamefont {Prigge}}, \bibinfo {author} {\bibfnamefont {G.~D.}\
  \bibnamefont {Cole}}, \bibinfo {author} {\bibfnamefont {M.}~\bibnamefont
  {Aspelmeyer}},\ and\ \bibinfo {author} {\bibfnamefont {J.}~\bibnamefont
  {Eisert}},\ }\bibfield  {title} {\bibinfo {title} {Observation of
  non-{M}arkovian micromechanical {B}rownian motion},\ }\href
  {https://doi.org/10.1038/ncomms8606} {\bibfield  {journal} {\bibinfo
  {journal} {Nat. Commun.}\ }\textbf {\bibinfo {volume} {6}},\ \bibinfo {pages}
  {7606} (\bibinfo {year} {2015})}\BibitemShut {NoStop}%
\bibitem [{\citenamefont {Dai}\ \emph {et~al.}(2025)\citenamefont {Dai},
  \citenamefont {Trappen}, \citenamefont {Chen}, \citenamefont {Melanson},
  \citenamefont {Yurtalan}, \citenamefont {Tennant}, \citenamefont {Martinez},
  \citenamefont {Tang}, \citenamefont {Mozgunov}, \citenamefont {Gibson},
  \citenamefont {Grover}, \citenamefont {Disseler}, \citenamefont {Basham},
  \citenamefont {Novikov}, \citenamefont {Das}, \citenamefont {Melville},
  \citenamefont {Niedzielski}, \citenamefont {Hirjibehedin}, \citenamefont
  {Serniak}, \citenamefont {Weber}, \citenamefont {Yoder}, \citenamefont
  {Oliver}, \citenamefont {Zick}, \citenamefont {Lidar},\ and\ \citenamefont
  {Lupascu}}]{Dai2025Jan}%
  \BibitemOpen
  \bibfield  {author} {\bibinfo {author} {\bibfnamefont {X.}~\bibnamefont
  {Dai}}, \bibinfo {author} {\bibfnamefont {R.}~\bibnamefont {Trappen}},
  \bibinfo {author} {\bibfnamefont {H.}~\bibnamefont {Chen}}, \bibinfo {author}
  {\bibfnamefont {D.}~\bibnamefont {Melanson}}, \bibinfo {author}
  {\bibfnamefont {M.~A.}\ \bibnamefont {Yurtalan}}, \bibinfo {author}
  {\bibfnamefont {D.~M.}\ \bibnamefont {Tennant}}, \bibinfo {author}
  {\bibfnamefont {A.~J.}\ \bibnamefont {Martinez}}, \bibinfo {author}
  {\bibfnamefont {Y.}~\bibnamefont {Tang}}, \bibinfo {author} {\bibfnamefont
  {E.}~\bibnamefont {Mozgunov}}, \bibinfo {author} {\bibfnamefont
  {J.}~\bibnamefont {Gibson}}, \bibinfo {author} {\bibfnamefont {J.~A.}\
  \bibnamefont {Grover}}, \bibinfo {author} {\bibfnamefont {S.~M.}\
  \bibnamefont {Disseler}}, \bibinfo {author} {\bibfnamefont {J.~I.}\
  \bibnamefont {Basham}}, \bibinfo {author} {\bibfnamefont {S.}~\bibnamefont
  {Novikov}}, \bibinfo {author} {\bibfnamefont {R.}~\bibnamefont {Das}},
  \bibinfo {author} {\bibfnamefont {A.~J.}\ \bibnamefont {Melville}}, \bibinfo
  {author} {\bibfnamefont {B.~M.}\ \bibnamefont {Niedzielski}}, \bibinfo
  {author} {\bibfnamefont {C.~F.}\ \bibnamefont {Hirjibehedin}}, \bibinfo
  {author} {\bibfnamefont {K.}~\bibnamefont {Serniak}}, \bibinfo {author}
  {\bibfnamefont {S.~J.}\ \bibnamefont {Weber}}, \bibinfo {author}
  {\bibfnamefont {J.~L.}\ \bibnamefont {Yoder}}, \bibinfo {author}
  {\bibfnamefont {W.~D.}\ \bibnamefont {Oliver}}, \bibinfo {author}
  {\bibfnamefont {K.~M.}\ \bibnamefont {Zick}}, \bibinfo {author}
  {\bibfnamefont {D.~A.}\ \bibnamefont {Lidar}},\ and\ \bibinfo {author}
  {\bibfnamefont {A.}~\bibnamefont {Lupascu}},\ }\bibfield  {title} {\bibinfo
  {title} {Dissipative {Landau-Zener} tunneling in the crossover regime from
  weak to strong environment coupling},\ }\href
  {https://doi.org/10.1038/s41467-024-55588-x} {\bibfield  {journal} {\bibinfo
  {journal} {Nat. Commun.}\ }\textbf {\bibinfo {volume} {16}},\ \bibinfo
  {pages} {329} (\bibinfo {year} {2025})}\BibitemShut {NoStop}%
\bibitem [{\citenamefont {Chin}\ \emph {et~al.}(2025)\citenamefont {Chin},
  \citenamefont {Keeling}, \citenamefont {Segal},\ and\ \citenamefont
  {Wang}}]{Chin2025Aug}%
  \BibitemOpen
  \bibfield  {author} {\bibinfo {author} {\bibfnamefont {A.}~\bibnamefont
  {Chin}}, \bibinfo {author} {\bibfnamefont {J.}~\bibnamefont {Keeling}},
  \bibinfo {author} {\bibfnamefont {D.}~\bibnamefont {Segal}},\ and\ \bibinfo
  {author} {\bibfnamefont {H.}~\bibnamefont {Wang}},\ }\bibfield  {title}
  {\bibinfo {title} {Algorithms and software for open quantum system
  dynamics},\ }\bibfield  {journal} {\bibinfo  {journal} {J. Chem. Phys.}\
  }\textbf {\bibinfo {volume} {163}},\ \href
  {https://doi.org/10.1063/5.0289390} {10.1063/5.0289390} (\bibinfo {year}
  {2025})\BibitemShut {NoStop}%
\bibitem [{\citenamefont {Cerrillo}\ and\ \citenamefont
  {Cao}(2014)}]{Cerrillo2014Mar}%
  \BibitemOpen
  \bibfield  {author} {\bibinfo {author} {\bibfnamefont {J.}~\bibnamefont
  {Cerrillo}}\ and\ \bibinfo {author} {\bibfnamefont {J.}~\bibnamefont {Cao}},\
  }\bibfield  {title} {\bibinfo {title} {Non-markovian dynamical maps:
  Numerical processing of open quantum trajectories},\ }\href
  {https://doi.org/10.1103/PhysRevLett.112.110401} {\bibfield  {journal}
  {\bibinfo  {journal} {Phys. Rev. Lett.}\ }\textbf {\bibinfo {volume} {112}},\
  \bibinfo {pages} {110401} (\bibinfo {year} {2014})}\BibitemShut {NoStop}%
\bibitem [{\citenamefont {Suess}\ \emph {et~al.}(2014)\citenamefont {Suess},
  \citenamefont {Eisfeld},\ and\ \citenamefont {Strunz}}]{HOPS}%
  \BibitemOpen
  \bibfield  {author} {\bibinfo {author} {\bibfnamefont {D.}~\bibnamefont
  {Suess}}, \bibinfo {author} {\bibfnamefont {A.}~\bibnamefont {Eisfeld}},\
  and\ \bibinfo {author} {\bibfnamefont {W.~T.}\ \bibnamefont {Strunz}},\
  }\bibfield  {title} {\bibinfo {title} {Hierarchy of stochastic pure states
  for open quantum system dynamics},\ }\href
  {https://doi.org/10.1103/PhysRevLett.113.150403} {\bibfield  {journal}
  {\bibinfo  {journal} {Phys. Rev. Lett.}\ }\textbf {\bibinfo {volume} {113}},\
  \bibinfo {pages} {150403} (\bibinfo {year} {2014})}\BibitemShut {NoStop}%
\bibitem [{\citenamefont {M{\ifmmode\ddot{u}\else\"{u}\fi}ller}\ and\
  \citenamefont {Strunz}(2025)}]{nuHOPS}%
  \BibitemOpen
  \bibfield  {author} {\bibinfo {author} {\bibfnamefont {K.}~\bibnamefont
  {M{\ifmmode\ddot{u}\else\"{u}\fi}ller}}\ and\ \bibinfo {author}
  {\bibfnamefont {W.~T.}\ \bibnamefont {Strunz}},\ }\bibfield  {title}
  {\bibinfo {title} {Quantum trajectory method for highly excited environments
  in non-{Markovian} open quantum dynamics},\ }\href
  {https://doi.org/10.1103/xhff-x24s} {\bibfield  {journal} {\bibinfo
  {journal} {Phys. Rev. A}\ }\textbf {\bibinfo {volume} {112}},\ \bibinfo
  {pages} {033719} (\bibinfo {year} {2025})}\BibitemShut {NoStop}%
\bibitem [{\citenamefont {Link}\ \emph {et~al.}(2023)\citenamefont {Link},
  \citenamefont {Luoma},\ and\ \citenamefont {Strunz}}]{Link_2023}%
  \BibitemOpen
  \bibfield  {author} {\bibinfo {author} {\bibfnamefont {V.}~\bibnamefont
  {Link}}, \bibinfo {author} {\bibfnamefont {K.}~\bibnamefont {Luoma}},\ and\
  \bibinfo {author} {\bibfnamefont {W.~T.}\ \bibnamefont {Strunz}},\ }\bibfield
   {title} {\bibinfo {title} {Non-markovian quantum state diffusion for spin
  environments},\ }\href {https://doi.org/10.1088/1367-2630/aceff3} {\bibfield
  {journal} {\bibinfo  {journal} {New Journal of Physics}\ }\textbf {\bibinfo
  {volume} {25}},\ \bibinfo {pages} {093006} (\bibinfo {year}
  {2023})}\BibitemShut {NoStop}%
\bibitem [{\citenamefont {Strathearn}\ \emph {et~al.}(2018)\citenamefont
  {Strathearn}, \citenamefont {Kirton}, \citenamefont {Kilda}, \citenamefont
  {Keeling},\ and\ \citenamefont {Lovett}}]{TEMPO}%
  \BibitemOpen
  \bibfield  {author} {\bibinfo {author} {\bibfnamefont {A.}~\bibnamefont
  {Strathearn}}, \bibinfo {author} {\bibfnamefont {P.}~\bibnamefont {Kirton}},
  \bibinfo {author} {\bibfnamefont {D.}~\bibnamefont {Kilda}}, \bibinfo
  {author} {\bibfnamefont {J.}~\bibnamefont {Keeling}},\ and\ \bibinfo {author}
  {\bibfnamefont {B.~W.}\ \bibnamefont {Lovett}},\ }\bibfield  {title}
  {\bibinfo {title} {Efficient non-markovian quantum dynamics using
  time-evolving matrix product operators},\ }\href
  {https://doi.org/10.1038/s41467-018-05617-3} {\bibfield  {journal} {\bibinfo
  {journal} {Nat. Commun.}\ }\textbf {\bibinfo {volume} {9}},\ \bibinfo {pages}
  {1} (\bibinfo {year} {2018})}\BibitemShut {NoStop}%
\bibitem [{\citenamefont {Fowler-Wright}\ \emph {et~al.}(2022)\citenamefont
  {Fowler-Wright}, \citenamefont {Lovett},\ and\ \citenamefont
  {Keeling}}]{Fowlwer-Wrigth_2022}%
  \BibitemOpen
  \bibfield  {author} {\bibinfo {author} {\bibfnamefont {P.}~\bibnamefont
  {Fowler-Wright}}, \bibinfo {author} {\bibfnamefont {B.~W.}\ \bibnamefont
  {Lovett}},\ and\ \bibinfo {author} {\bibfnamefont {J.}~\bibnamefont
  {Keeling}},\ }\bibfield  {title} {\bibinfo {title} {Efficient many-body
  non-markovian dynamics of organic polaritons},\ }\href
  {https://doi.org/10.1103/PhysRevLett.129.173001} {\bibfield  {journal}
  {\bibinfo  {journal} {Phys. Rev. Lett.}\ }\textbf {\bibinfo {volume} {129}},\
  \bibinfo {pages} {173001} (\bibinfo {year} {2022})}\BibitemShut {NoStop}%
\bibitem [{\citenamefont {Link}\ \emph {et~al.}(2024)\citenamefont {Link},
  \citenamefont {Tu},\ and\ \citenamefont {Strunz}}]{uniTEMPO}%
  \BibitemOpen
  \bibfield  {author} {\bibinfo {author} {\bibfnamefont {V.}~\bibnamefont
  {Link}}, \bibinfo {author} {\bibfnamefont {H.-H.}\ \bibnamefont {Tu}},\ and\
  \bibinfo {author} {\bibfnamefont {W.~T.}\ \bibnamefont {Strunz}},\ }\bibfield
   {title} {\bibinfo {title} {Open quantum system dynamics from infinite tensor
  network contraction},\ }\href
  {https://doi.org/10.1103/PhysRevLett.132.200403} {\bibfield  {journal}
  {\bibinfo  {journal} {Phys. Rev. Lett.}\ }\textbf {\bibinfo {volume} {132}},\
  \bibinfo {pages} {200403} (\bibinfo {year} {2024})}\BibitemShut {NoStop}%
\bibitem [{\citenamefont {Cygorek}\ \emph {et~al.}(2022)\citenamefont
  {Cygorek}, \citenamefont {Cosacchi}, \citenamefont {Vagov}, \citenamefont
  {Axt}, \citenamefont {Lovett}, \citenamefont {Keeling},\ and\ \citenamefont
  {Gauger}}]{ACE}%
  \BibitemOpen
  \bibfield  {author} {\bibinfo {author} {\bibfnamefont {M.}~\bibnamefont
  {Cygorek}}, \bibinfo {author} {\bibfnamefont {M.}~\bibnamefont {Cosacchi}},
  \bibinfo {author} {\bibfnamefont {A.}~\bibnamefont {Vagov}}, \bibinfo
  {author} {\bibfnamefont {V.~M.}\ \bibnamefont {Axt}}, \bibinfo {author}
  {\bibfnamefont {B.~W.}\ \bibnamefont {Lovett}}, \bibinfo {author}
  {\bibfnamefont {J.}~\bibnamefont {Keeling}},\ and\ \bibinfo {author}
  {\bibfnamefont {E.~M.}\ \bibnamefont {Gauger}},\ }\bibfield  {title}
  {\bibinfo {title} {Simulation of open quantum systems by automated
  compression of arbitrary environments},\ }\href
  {https://doi.org/10.1038/s41567-022-01544-9} {\bibfield  {journal} {\bibinfo
  {journal} {Nat. Phys.}\ }\textbf {\bibinfo {volume} {18}},\ \bibinfo {pages}
  {662} (\bibinfo {year} {2022})}\BibitemShut {NoStop}%
\bibitem [{\citenamefont {Prior}\ \emph {et~al.}(2010)\citenamefont {Prior},
  \citenamefont {Chin}, \citenamefont {Huelga},\ and\ \citenamefont
  {Plenio}}]{TEDOPA}%
  \BibitemOpen
  \bibfield  {author} {\bibinfo {author} {\bibfnamefont {J.}~\bibnamefont
  {Prior}}, \bibinfo {author} {\bibfnamefont {A.~W.}\ \bibnamefont {Chin}},
  \bibinfo {author} {\bibfnamefont {S.~F.}\ \bibnamefont {Huelga}},\ and\
  \bibinfo {author} {\bibfnamefont {M.~B.}\ \bibnamefont {Plenio}},\ }\bibfield
   {title} {\bibinfo {title} {Efficient simulation of strong system-environment
  interactions},\ }\href {https://doi.org/10.1103/PhysRevLett.105.050404}
  {\bibfield  {journal} {\bibinfo  {journal} {Phys. Rev. Lett.}\ }\textbf
  {\bibinfo {volume} {105}},\ \bibinfo {pages} {050404} (\bibinfo {year}
  {2010})}\BibitemShut {NoStop}%
\bibitem [{\citenamefont {Lacroix}\ \emph {et~al.}(2024)\citenamefont
  {Lacroix}, \citenamefont {Le~D{\ifmmode\acute{e}\else\'{e}\fi}},
  \citenamefont {Riva}, \citenamefont {Dunnett},\ and\ \citenamefont
  {Chin}}]{TEDOPA_algorithm}%
  \BibitemOpen
  \bibfield  {author} {\bibinfo {author} {\bibfnamefont {T.}~\bibnamefont
  {Lacroix}}, \bibinfo {author} {\bibfnamefont {B.}~\bibnamefont
  {Le~D{\ifmmode\acute{e}\else\'{e}\fi}}}, \bibinfo {author} {\bibfnamefont
  {A.}~\bibnamefont {Riva}}, \bibinfo {author} {\bibfnamefont {A.~J.}\
  \bibnamefont {Dunnett}},\ and\ \bibinfo {author} {\bibfnamefont {A.~W.}\
  \bibnamefont {Chin}},\ }\bibfield  {title} {\bibinfo {title}
  {{MPSDynamics.jl: Tensor network simulations for finite-temperature
  (non-Markovian) open quantum system dynamics}},\ }\href
  {https://doi.org/10.1063/5.0223107} {\bibfield  {journal} {\bibinfo
  {journal} {J. Chem. Phys.}\ }\textbf {\bibinfo {volume} {161}},\ \bibinfo
  {pages} {084116} (\bibinfo {year} {2024})}\BibitemShut {NoStop}%
\bibitem [{\citenamefont {Garraway}(1997)}]{Garraway1997Mar}%
  \BibitemOpen
  \bibfield  {author} {\bibinfo {author} {\bibfnamefont {B.~M.}\ \bibnamefont
  {Garraway}},\ }\bibfield  {title} {\bibinfo {title} {Nonperturbative decay of
  an atomic system in a cavity},\ }\href
  {https://doi.org/10.1103/PhysRevA.55.2290} {\bibfield  {journal} {\bibinfo
  {journal} {Phys. Rev. A}\ }\textbf {\bibinfo {volume} {55}},\ \bibinfo
  {pages} {2290} (\bibinfo {year} {1997})}\BibitemShut {NoStop}%
\bibitem [{\citenamefont {Pleasance}\ \emph {et~al.}(2020)\citenamefont
  {Pleasance}, \citenamefont {Garraway},\ and\ \citenamefont
  {Petruccione}}]{Pleasance2020Oct}%
  \BibitemOpen
  \bibfield  {author} {\bibinfo {author} {\bibfnamefont {G.}~\bibnamefont
  {Pleasance}}, \bibinfo {author} {\bibfnamefont {B.~M.}\ \bibnamefont
  {Garraway}},\ and\ \bibinfo {author} {\bibfnamefont {F.}~\bibnamefont
  {Petruccione}},\ }\bibfield  {title} {\bibinfo {title} {Generalized theory of
  pseudomodes for exact descriptions of non-markovian quantum processes},\
  }\href {https://doi.org/10.1103/PhysRevResearch.2.043058} {\bibfield
  {journal} {\bibinfo  {journal} {Phys. Rev. Res.}\ }\textbf {\bibinfo {volume}
  {2}},\ \bibinfo {pages} {043058} (\bibinfo {year} {2020})}\BibitemShut
  {NoStop}%
\bibitem [{\citenamefont {Mascherpa}\ \emph {et~al.}(2020)\citenamefont
  {Mascherpa}, \citenamefont {Smirne}, \citenamefont {Somoza}, \citenamefont
  {Fern\'andez-Acebal}, \citenamefont {Donadi}, \citenamefont {Tamascelli},
  \citenamefont {Huelga},\ and\ \citenamefont {Plenio}}]{Pseudomode_Plenio}%
  \BibitemOpen
  \bibfield  {author} {\bibinfo {author} {\bibfnamefont {F.}~\bibnamefont
  {Mascherpa}}, \bibinfo {author} {\bibfnamefont {A.}~\bibnamefont {Smirne}},
  \bibinfo {author} {\bibfnamefont {A.~D.}\ \bibnamefont {Somoza}}, \bibinfo
  {author} {\bibfnamefont {P.}~\bibnamefont {Fern\'andez-Acebal}}, \bibinfo
  {author} {\bibfnamefont {S.}~\bibnamefont {Donadi}}, \bibinfo {author}
  {\bibfnamefont {D.}~\bibnamefont {Tamascelli}}, \bibinfo {author}
  {\bibfnamefont {S.~F.}\ \bibnamefont {Huelga}},\ and\ \bibinfo {author}
  {\bibfnamefont {M.~B.}\ \bibnamefont {Plenio}},\ }\bibfield  {title}
  {\bibinfo {title} {Optimized auxiliary oscillators for the simulation of
  general open quantum systems},\ }\href
  {https://doi.org/10.1103/PhysRevA.101.052108} {\bibfield  {journal} {\bibinfo
   {journal} {Phys. Rev. A}\ }\textbf {\bibinfo {volume} {101}},\ \bibinfo
  {pages} {052108} (\bibinfo {year} {2020})}\BibitemShut {NoStop}%
\bibitem [{\citenamefont {Huang}\ \emph {et~al.}(2025)\citenamefont {Huang},
  \citenamefont {Park}, \citenamefont {Chan},\ and\ \citenamefont
  {Lin}}]{pseudomodes_lin}%
  \BibitemOpen
  \bibfield  {author} {\bibinfo {author} {\bibfnamefont {Z.}~\bibnamefont
  {Huang}}, \bibinfo {author} {\bibfnamefont {G.}~\bibnamefont {Park}},
  \bibinfo {author} {\bibfnamefont {G.~K.-L.}\ \bibnamefont {Chan}},\ and\
  \bibinfo {author} {\bibfnamefont {L.}~\bibnamefont {Lin}},\ }\bibfield
  {title} {\bibinfo {title} {Coupled lindblad pseudomode theory for simulating
  open quantum systems},\ }\bibfield  {journal} {\bibinfo  {journal} {arXiv}\
  }\href {https://doi.org/10.48550/arXiv.2506.10308}
  {10.48550/arXiv.2506.10308} (\bibinfo {year} {2025}),\ \Eprint
  {https://arxiv.org/abs/2506.10308} {2506.10308} \BibitemShut {NoStop}%
\bibitem [{\citenamefont {Medina}\ \emph {et~al.}(2021)\citenamefont {Medina},
  \citenamefont {Garc\'{\i}a-Vidal}, \citenamefont
  {Fern\'andez-Dom\'{\i}nguez},\ and\ \citenamefont
  {Feist}}]{pseudomodes_Feist}%
  \BibitemOpen
  \bibfield  {author} {\bibinfo {author} {\bibfnamefont {I.}~\bibnamefont
  {Medina}}, \bibinfo {author} {\bibfnamefont {F.~J.}\ \bibnamefont
  {Garc\'{\i}a-Vidal}}, \bibinfo {author} {\bibfnamefont {A.~I.}\ \bibnamefont
  {Fern\'andez-Dom\'{\i}nguez}},\ and\ \bibinfo {author} {\bibfnamefont
  {J.}~\bibnamefont {Feist}},\ }\bibfield  {title} {\bibinfo {title} {Few-mode
  field quantization of arbitrary electromagnetic spectral densities},\ }\href
  {https://doi.org/10.1103/PhysRevLett.126.093601} {\bibfield  {journal}
  {\bibinfo  {journal} {Phys. Rev. Lett.}\ }\textbf {\bibinfo {volume} {126}},\
  \bibinfo {pages} {093601} (\bibinfo {year} {2021})}\BibitemShut {NoStop}%
\bibitem [{\citenamefont {Wang}\ and\ \citenamefont {Thoss}(2003)}]{ML-MCTDH}%
  \BibitemOpen
  \bibfield  {author} {\bibinfo {author} {\bibfnamefont {H.}~\bibnamefont
  {Wang}}\ and\ \bibinfo {author} {\bibfnamefont {M.}~\bibnamefont {Thoss}},\
  }\bibfield  {title} {\bibinfo {title} {Multilayer formulation of the
  multiconfiguration time-dependent {Hartree} theory},\ }\href
  {https://doi.org/10.1063/1.1580111} {\bibfield  {journal} {\bibinfo
  {journal} {J. Chem. Phys.}\ }\textbf {\bibinfo {volume} {119}},\ \bibinfo
  {pages} {1289} (\bibinfo {year} {2003})}\BibitemShut {NoStop}%
\bibitem [{\citenamefont {Carmichael}(1999)}]{carmichael1999statistical}%
  \BibitemOpen
  \bibfield  {author} {\bibinfo {author} {\bibfnamefont {H.}~\bibnamefont
  {Carmichael}},\ }\href {https://books.google.fi/books?id=GcDvAAAAMAAJ} {\emph
  {\bibinfo {title} {Statistical Methods in Quantum Optics 1: Master Equations
  and Fokker-Planck Equations}}},\ Physics and astronomy online library\
  (\bibinfo  {publisher} {Springer},\ \bibinfo {year} {1999})\BibitemShut
  {NoStop}%
\bibitem [{\citenamefont {Zhao}(2023)}]{Davydov_review}%
  \BibitemOpen
  \bibfield  {author} {\bibinfo {author} {\bibfnamefont {Y.}~\bibnamefont
  {Zhao}},\ }\bibfield  {title} {\bibinfo {title} {The hierarchy of
  {D}avydov{'}s ans{\ifmmode\ddot{a}\else\"{a}\fi}tze: {F}rom guesswork to
  numerically {\textquotedblleft}exact{\textquotedblright} many-body wave
  functions},\ }\bibfield  {journal} {\bibinfo  {journal} {J. Chem. Phys.}\
  }\textbf {\bibinfo {volume} {158}},\ \href
  {https://doi.org/10.1063/5.0140002} {10.1063/5.0140002} (\bibinfo {year}
  {2023})\BibitemShut {NoStop}%
\bibitem [{\citenamefont {Werther}\ and\ \citenamefont
  {Gro{\ss}mann}(2020)}]{Davydov_Frank}%
  \BibitemOpen
  \bibfield  {author} {\bibinfo {author} {\bibfnamefont {M.}~\bibnamefont
  {Werther}}\ and\ \bibinfo {author} {\bibfnamefont {F.}~\bibnamefont
  {Gro{\ss}mann}},\ }\bibfield  {title} {\bibinfo {title} {Apoptosis of moving
  nonorthogonal basis functions in many-particle quantum dynamics},\ }\href
  {https://doi.org/10.1103/PhysRevB.101.174315} {\bibfield  {journal} {\bibinfo
   {journal} {Phys. Rev. B}\ }\textbf {\bibinfo {volume} {101}},\ \bibinfo
  {pages} {174315} (\bibinfo {year} {2020})}\BibitemShut {NoStop}%
\bibitem [{\citenamefont {Tanimura}(1990)}]{Tanimura_1990}%
  \BibitemOpen
  \bibfield  {author} {\bibinfo {author} {\bibfnamefont {Y.}~\bibnamefont
  {Tanimura}},\ }\bibfield  {title} {\bibinfo {title} {Nonperturbative
  expansion method for a quantum system coupled to a harmonic-oscillator
  bath},\ }\href {https://doi.org/10.1103/PhysRevA.41.6676} {\bibfield
  {journal} {\bibinfo  {journal} {Phys. Rev. A}\ }\textbf {\bibinfo {volume}
  {41}},\ \bibinfo {pages} {6676} (\bibinfo {year} {1990})}\BibitemShut
  {NoStop}%
\bibitem [{\citenamefont {Tanimura}(2020)}]{Tanimura2020}%
  \BibitemOpen
  \bibfield  {author} {\bibinfo {author} {\bibfnamefont {Y.}~\bibnamefont
  {Tanimura}},\ }\bibfield  {title} {\bibinfo {title} {{Numerically “exact”
  approach to open quantum dynamics: The hierarchical equations of motion
  (HEOM)}},\ }\href {https://doi.org/10.1063/5.0011599} {\bibfield  {journal}
  {\bibinfo  {journal} {The Journal of Chemical Physics}\ }\textbf {\bibinfo
  {volume} {153}},\ \bibinfo {pages} {020901} (\bibinfo {year}
  {2020})}\BibitemShut {NoStop}%
\bibitem [{\citenamefont {Feynman}\ and\ \citenamefont
  {Vernon}(1963)}]{Feynman1963Oct}%
  \BibitemOpen
  \bibfield  {author} {\bibinfo {author} {\bibfnamefont {R.~P.}\ \bibnamefont
  {Feynman}}\ and\ \bibinfo {author} {\bibfnamefont {F.~L.}\ \bibnamefont
  {Vernon}},\ }\bibfield  {title} {\bibinfo {title} {The theory of a general
  quantum system interacting with a linear dissipative system},\ }\href
  {https://doi.org/10.1016/0003-4916(63)90068-X} {\bibfield  {journal}
  {\bibinfo  {journal} {Ann. Phys.}\ }\textbf {\bibinfo {volume} {24}},\
  \bibinfo {pages} {118} (\bibinfo {year} {1963})}\BibitemShut {NoStop}%
\bibitem [{\citenamefont
  {Imamog\ifmmode\bar\else\textasciimacron\fi{}lu}(1994)}]{pseudomodes_Imamoglu}%
  \BibitemOpen
  \bibfield  {author} {\bibinfo {author} {\bibfnamefont {A.}~\bibnamefont
  {Imamog\ifmmode\bar\else\textasciimacron\fi{}lu}},\ }\bibfield  {title}
  {\bibinfo {title} {Stochastic wave-function approach to non-markovian
  systems},\ }\href {https://doi.org/10.1103/PhysRevA.50.3650} {\bibfield
  {journal} {\bibinfo  {journal} {Phys. Rev. A}\ }\textbf {\bibinfo {volume}
  {50}},\ \bibinfo {pages} {3650} (\bibinfo {year} {1994})}\BibitemShut
  {NoStop}%
\bibitem [{\citenamefont {Zhou}\ \emph {et~al.}(2024)\citenamefont {Zhou},
  \citenamefont {Jin},\ and\ \citenamefont {Yang}}]{pseudomodes_Yang}%
  \BibitemOpen
  \bibfield  {author} {\bibinfo {author} {\bibfnamefont {L.~K.}\ \bibnamefont
  {Zhou}}, \bibinfo {author} {\bibfnamefont {G.~R.}\ \bibnamefont {Jin}},\ and\
  \bibinfo {author} {\bibfnamefont {W.}~\bibnamefont {Yang}},\ }\bibfield
  {title} {\bibinfo {title} {Systematic and efficient pseudomode method to
  simulate open quantum systems under a bosonic environment},\ }\href
  {https://doi.org/10.1103/PhysRevA.110.022221} {\bibfield  {journal} {\bibinfo
   {journal} {Phys. Rev. A}\ }\textbf {\bibinfo {volume} {110}},\ \bibinfo
  {pages} {022221} (\bibinfo {year} {2024})}\BibitemShut {NoStop}%
\bibitem [{\citenamefont {Hughes}\ \emph {et~al.}(2018)\citenamefont {Hughes},
  \citenamefont {Richter},\ and\ \citenamefont {Knorr}}]{pseudomodes_Knorr}%
  \BibitemOpen
  \bibfield  {author} {\bibinfo {author} {\bibfnamefont {S.}~\bibnamefont
  {Hughes}}, \bibinfo {author} {\bibfnamefont {M.}~\bibnamefont {Richter}},\
  and\ \bibinfo {author} {\bibfnamefont {A.}~\bibnamefont {Knorr}},\ }\bibfield
   {title} {\bibinfo {title} {Quantized pseudomodes for plasmonic cavity
  {QED}},\ }\href {https://doi.org/10.1364/OL.43.001834} {\bibfield  {journal}
  {\bibinfo  {journal} {Opt. Lett.}\ }\textbf {\bibinfo {volume} {43}},\
  \bibinfo {pages} {1834} (\bibinfo {year} {2018})}\BibitemShut {NoStop}%
\bibitem [{\citenamefont {Lentrodt}\ and\ \citenamefont
  {Evers}(2020)}]{Lentrodt20}%
  \BibitemOpen
  \bibfield  {author} {\bibinfo {author} {\bibfnamefont {D.}~\bibnamefont
  {Lentrodt}}\ and\ \bibinfo {author} {\bibfnamefont {J.}~\bibnamefont
  {Evers}},\ }\bibfield  {title} {\bibinfo {title} {Ab initio few-mode theory
  for quantum potential scattering problems},\ }\href
  {https://doi.org/10.1103/PhysRevX.10.011008} {\bibfield  {journal} {\bibinfo
  {journal} {Phys. Rev. X}\ }\textbf {\bibinfo {volume} {10}},\ \bibinfo
  {pages} {011008} (\bibinfo {year} {2020})}\BibitemShut {NoStop}%
\bibitem [{\citenamefont {Bulla}\ \emph {et~al.}(2005)\citenamefont {Bulla},
  \citenamefont {Lee}, \citenamefont {Tong},\ and\ \citenamefont
  {Vojta}}]{Chain_Voijta}%
  \BibitemOpen
  \bibfield  {author} {\bibinfo {author} {\bibfnamefont {R.}~\bibnamefont
  {Bulla}}, \bibinfo {author} {\bibfnamefont {H.-J.}\ \bibnamefont {Lee}},
  \bibinfo {author} {\bibfnamefont {N.-H.}\ \bibnamefont {Tong}},\ and\
  \bibinfo {author} {\bibfnamefont {M.}~\bibnamefont {Vojta}},\ }\bibfield
  {title} {\bibinfo {title} {Numerical renormalization group for quantum
  impurities in a bosonic bath},\ }\href
  {https://doi.org/10.1103/PhysRevB.71.045122} {\bibfield  {journal} {\bibinfo
  {journal} {Phys. Rev. B}\ }\textbf {\bibinfo {volume} {71}},\ \bibinfo
  {pages} {045122} (\bibinfo {year} {2005})}\BibitemShut {NoStop}%
\bibitem [{\citenamefont {Hughes}\ \emph {et~al.}(2009)\citenamefont {Hughes},
  \citenamefont {Christ},\ and\ \citenamefont {Burghardt}}]{Chain_Burghardt}%
  \BibitemOpen
  \bibfield  {author} {\bibinfo {author} {\bibfnamefont {K.~H.}\ \bibnamefont
  {Hughes}}, \bibinfo {author} {\bibfnamefont {C.~D.}\ \bibnamefont {Christ}},\
  and\ \bibinfo {author} {\bibfnamefont {I.}~\bibnamefont {Burghardt}},\
  }\bibfield  {title} {\bibinfo {title} {{Effective-mode representation of
  non-Markovian dynamics: A hierarchical approximation of the spectral density.
  I. Application to single surface dynamics}},\ }\href
  {https://doi.org/10.1063/1.3159671} {\bibfield  {journal} {\bibinfo
  {journal} {J. Chem. Phys.}\ }\textbf {\bibinfo {volume} {131}},\ \bibinfo
  {pages} {024109} (\bibinfo {year} {2009})}\BibitemShut {NoStop}%
\bibitem [{\citenamefont {Chin}\ \emph {et~al.}(2010)\citenamefont {Chin},
  \citenamefont {Rivas}, \citenamefont {Huelga},\ and\ \citenamefont
  {Plenio}}]{Chain_Plenio}%
  \BibitemOpen
  \bibfield  {author} {\bibinfo {author} {\bibfnamefont {A.~W.}\ \bibnamefont
  {Chin}}, \bibinfo {author} {\bibfnamefont
  {{\ifmmode\acute{A}\else\'{A}\fi}.}~\bibnamefont {Rivas}}, \bibinfo {author}
  {\bibfnamefont {S.~F.}\ \bibnamefont {Huelga}},\ and\ \bibinfo {author}
  {\bibfnamefont {M.~B.}\ \bibnamefont {Plenio}},\ }\bibfield  {title}
  {\bibinfo {title} {Exact mapping between system-reservoir quantum models and
  semi-infinite discrete chains using orthogonal polynomials},\ }\href
  {https://doi.org/10.1063/1.3490188} {\bibfield  {journal} {\bibinfo
  {journal} {J. Math. Phys.}\ }\textbf {\bibinfo {volume} {51}},\ \bibinfo
  {pages} {092109} (\bibinfo {year} {2010})}\BibitemShut {NoStop}%
\bibitem [{\citenamefont {S{\ifmmode\acute{a}\else\'{a}\fi}nchez-Barquilla}\
  and\ \citenamefont {Feist}(2021)}]{Chain_Feist}%
  \BibitemOpen
  \bibfield  {author} {\bibinfo {author} {\bibfnamefont {M.}~\bibnamefont
  {S{\ifmmode\acute{a}\else\'{a}\fi}nchez-Barquilla}}\ and\ \bibinfo {author}
  {\bibfnamefont {J.}~\bibnamefont {Feist}},\ }\bibfield  {title} {\bibinfo
  {title} {Accurate truncations of chain mapping models for open quantum
  systems},\ }\href {https://doi.org/10.3390/nano11082104} {\bibfield
  {journal} {\bibinfo  {journal} {Nanomaterials}\ }\textbf {\bibinfo {volume}
  {11}},\ \bibinfo {pages} {2104} (\bibinfo {year} {2021})}\BibitemShut
  {NoStop}%
\bibitem [{\citenamefont {Hsieh}\ and\ \citenamefont
  {Cao}(2018)}]{Hsieh2018Jan}%
  \BibitemOpen
  \bibfield  {author} {\bibinfo {author} {\bibfnamefont {C.-Y.}\ \bibnamefont
  {Hsieh}}\ and\ \bibinfo {author} {\bibfnamefont {J.}~\bibnamefont {Cao}},\
  }\bibfield  {title} {\bibinfo {title} {A unified stochastic formulation of
  dissipative quantum dynamics. {I. Generalized hierarchical equations}},\
  }\bibfield  {journal} {\bibinfo  {journal} {J. Chem. Phys.}\ }\textbf
  {\bibinfo {volume} {148}},\ \href {https://doi.org/10.1063/1.5018725}
  {10.1063/1.5018725} (\bibinfo {year} {2018})\BibitemShut {NoStop}%
\bibitem [{\citenamefont {Jin}\ \emph {et~al.}(2008)\citenamefont {Jin},
  \citenamefont {Zheng},\ and\ \citenamefont {Yan}}]{Jin2008Jun}%
  \BibitemOpen
  \bibfield  {author} {\bibinfo {author} {\bibfnamefont {J.}~\bibnamefont
  {Jin}}, \bibinfo {author} {\bibfnamefont {X.}~\bibnamefont {Zheng}},\ and\
  \bibinfo {author} {\bibfnamefont {Y.}~\bibnamefont {Yan}},\ }\bibfield
  {title} {\bibinfo {title} {Exact dynamics of dissipative electronic systems
  and quantum transport: {H}ierarchical equations of motion approach},\ }\href
  {https://doi.org/10.1063/1.2938087} {\bibfield  {journal} {\bibinfo
  {journal} {J. Chem. Phys.}\ }\textbf {\bibinfo {volume} {128}},\ \bibinfo
  {pages} {234703} (\bibinfo {year} {2008})}\BibitemShut {NoStop}%
\bibitem [{\citenamefont {B{\ifmmode\ddot{a}\else\"{a}\fi}tge}\ \emph
  {et~al.}(2021)\citenamefont {B{\ifmmode\ddot{a}\else\"{a}\fi}tge},
  \citenamefont {Ke}, \citenamefont {Kaspar},\ and\ \citenamefont
  {Thoss}}]{Batge2021Jun}%
  \BibitemOpen
  \bibfield  {author} {\bibinfo {author} {\bibfnamefont {J.}~\bibnamefont
  {B{\ifmmode\ddot{a}\else\"{a}\fi}tge}}, \bibinfo {author} {\bibfnamefont
  {Y.}~\bibnamefont {Ke}}, \bibinfo {author} {\bibfnamefont {C.}~\bibnamefont
  {Kaspar}},\ and\ \bibinfo {author} {\bibfnamefont {M.}~\bibnamefont
  {Thoss}},\ }\bibfield  {title} {\bibinfo {title} {Nonequilibrium open quantum
  systems with multiple bosonic and fermionic environments: {A hierarchical
  equations of motion approach}},\ }\href
  {https://doi.org/10.1103/PhysRevB.103.235413} {\bibfield  {journal} {\bibinfo
   {journal} {Phys. Rev. B}\ }\textbf {\bibinfo {volume} {103}},\ \bibinfo
  {pages} {235413} (\bibinfo {year} {2021})}\BibitemShut {NoStop}%
\bibitem [{\citenamefont {Debecker}\ \emph {et~al.}(2024)\citenamefont
  {Debecker}, \citenamefont {Martin},\ and\ \citenamefont
  {Damanet}}]{Debecker2024Oct}%
  \BibitemOpen
  \bibfield  {author} {\bibinfo {author} {\bibfnamefont {B.}~\bibnamefont
  {Debecker}}, \bibinfo {author} {\bibfnamefont {J.}~\bibnamefont {Martin}},\
  and\ \bibinfo {author} {\bibfnamefont {F.}~\bibnamefont {Damanet}},\
  }\bibfield  {title} {\bibinfo {title} {Spectral theory of non-{M}arkovian
  dissipative phase transitions},\ }\href
  {https://doi.org/10.1103/PhysRevA.110.042201} {\bibfield  {journal} {\bibinfo
   {journal} {Phys. Rev. A}\ }\textbf {\bibinfo {volume} {110}},\ \bibinfo
  {pages} {042201} (\bibinfo {year} {2024})}\BibitemShut {NoStop}%
\bibitem [{\citenamefont {Xu}\ \emph {et~al.}(2022)\citenamefont {Xu},
  \citenamefont {Yan}, \citenamefont {Shi}, \citenamefont {Ankerhold},\ and\
  \citenamefont {Stockburger}}]{Xu2022Nov}%
  \BibitemOpen
  \bibfield  {author} {\bibinfo {author} {\bibfnamefont {M.}~\bibnamefont
  {Xu}}, \bibinfo {author} {\bibfnamefont {Y.}~\bibnamefont {Yan}}, \bibinfo
  {author} {\bibfnamefont {Q.}~\bibnamefont {Shi}}, \bibinfo {author}
  {\bibfnamefont {J.}~\bibnamefont {Ankerhold}},\ and\ \bibinfo {author}
  {\bibfnamefont {J.~T.}\ \bibnamefont {Stockburger}},\ }\bibfield  {title}
  {\bibinfo {title} {Taming quantum noise for efficient low temperature
  simulations of open quantum systems},\ }\href
  {https://doi.org/10.1103/PhysRevLett.129.230601} {\bibfield  {journal}
  {\bibinfo  {journal} {Phys. Rev. Lett.}\ }\textbf {\bibinfo {volume} {129}},\
  \bibinfo {pages} {230601} (\bibinfo {year} {2022})}\BibitemShut {NoStop}%
\bibitem [{\citenamefont {Hartmann}\ and\ \citenamefont
  {Strunz}(2017)}]{HOPS_Richard}%
  \BibitemOpen
  \bibfield  {author} {\bibinfo {author} {\bibfnamefont {R.}~\bibnamefont
  {Hartmann}}\ and\ \bibinfo {author} {\bibfnamefont {W.~T.}\ \bibnamefont
  {Strunz}},\ }\bibfield  {title} {\bibinfo {title} {Exact open quantum system
  dynamics using the hierarchy of pure states ({HOPS})},\ }\href
  {https://doi.org/10.1021/acs.jctc.7b00751} {\bibfield  {journal} {\bibinfo
  {journal} {Journal of Chemical Theory and Computation}\ }\textbf {\bibinfo
  {volume} {13}},\ \bibinfo {pages} {5834} (\bibinfo {year}
  {2017})}\BibitemShut {NoStop}%
\bibitem [{\citenamefont {Citty}\ \emph {et~al.}(2024)\citenamefont {Citty},
  \citenamefont {Lynd}, \citenamefont {Gera}, \citenamefont {Varvelo},\ and\
  \citenamefont {Raccah}}]{mesoHOPS}%
  \BibitemOpen
  \bibfield  {author} {\bibinfo {author} {\bibfnamefont {B.}~\bibnamefont
  {Citty}}, \bibinfo {author} {\bibfnamefont {J.~K.}\ \bibnamefont {Lynd}},
  \bibinfo {author} {\bibfnamefont {T.}~\bibnamefont {Gera}}, \bibinfo {author}
  {\bibfnamefont {L.}~\bibnamefont {Varvelo}},\ and\ \bibinfo {author}
  {\bibfnamefont {D.~I. G.~B.}\ \bibnamefont {Raccah}},\ }\bibfield  {title}
  {\bibinfo {title} {{MesoHOPS}: Size-invariant scaling calculations of
  multi-excitation open quantum systems},\ }\bibfield  {journal} {\bibinfo
  {journal} {J. Chem. Phys.}\ }\textbf {\bibinfo {volume} {160}},\ \href
  {https://doi.org/10.1063/5.0197825} {10.1063/5.0197825} (\bibinfo {year}
  {2024})\BibitemShut {NoStop}%
\bibitem [{\citenamefont {Takahashi}\ \emph {et~al.}(2024)\citenamefont
  {Takahashi}, \citenamefont {Rudge}, \citenamefont {Kaspar}, \citenamefont
  {Thoss},\ and\ \citenamefont {Borrelli}}]{expFittingOverview}%
  \BibitemOpen
  \bibfield  {author} {\bibinfo {author} {\bibfnamefont {H.}~\bibnamefont
  {Takahashi}}, \bibinfo {author} {\bibfnamefont {S.}~\bibnamefont {Rudge}},
  \bibinfo {author} {\bibfnamefont {C.}~\bibnamefont {Kaspar}}, \bibinfo
  {author} {\bibfnamefont {M.}~\bibnamefont {Thoss}},\ and\ \bibinfo {author}
  {\bibfnamefont {R.}~\bibnamefont {Borrelli}},\ }\bibfield  {title} {\bibinfo
  {title} {High accuracy exponential decomposition of bath correlation
  functions for arbitrary and structured spectral densities: Emerging
  methodologies and new approaches},\ }\href
  {https://doi.org/10.1063/5.0209348} {\bibfield  {journal} {\bibinfo
  {journal} {J. Chem. Phys.}\ }\textbf {\bibinfo {volume} {160}},\ \bibinfo
  {pages} {204105} (\bibinfo {year} {2024})}\BibitemShut {NoStop}%
\bibitem [{\citenamefont {Dunn}\ \emph {et~al.}(2019)\citenamefont {Dunn},
  \citenamefont {Tempelaar},\ and\ \citenamefont {Reichman}}]{Dunn2019May}%
  \BibitemOpen
  \bibfield  {author} {\bibinfo {author} {\bibfnamefont {I.~S.}\ \bibnamefont
  {Dunn}}, \bibinfo {author} {\bibfnamefont {R.}~\bibnamefont {Tempelaar}},\
  and\ \bibinfo {author} {\bibfnamefont {D.~R.}\ \bibnamefont {Reichman}},\
  }\bibfield  {title} {\bibinfo {title} {Removing instabilities in the
  hierarchical equations of motion: {E}xact and approximate projection
  approaches},\ }\bibfield  {journal} {\bibinfo  {journal} {J. Chem. Phys.}\
  }\textbf {\bibinfo {volume} {150}},\ \href
  {https://doi.org/10.1063/1.5092616} {10.1063/1.5092616} (\bibinfo {year}
  {2019})\BibitemShut {NoStop}%
\bibitem [{\citenamefont {Yan}\ \emph {et~al.}(2020)\citenamefont {Yan},
  \citenamefont {Xing},\ and\ \citenamefont {Shi}}]{Yan2020Nov}%
  \BibitemOpen
  \bibfield  {author} {\bibinfo {author} {\bibfnamefont {Y.}~\bibnamefont
  {Yan}}, \bibinfo {author} {\bibfnamefont {T.}~\bibnamefont {Xing}},\ and\
  \bibinfo {author} {\bibfnamefont {Q.}~\bibnamefont {Shi}},\ }\bibfield
  {title} {\bibinfo {title} {A new method to improve the numerical stability of
  the hierarchical equations of motion for discrete harmonic oscillator
  modes},\ }\href {https://doi.org/10.1063/5.0027962} {\bibfield  {journal}
  {\bibinfo  {journal} {J. Chem. Phys.}\ }\textbf {\bibinfo {volume} {153}},\
  \bibinfo {pages} {204109} (\bibinfo {year} {2020})}\BibitemShut {NoStop}%
\bibitem [{\citenamefont {Krug}\ and\ \citenamefont
  {Stockburger}(2023)}]{Krug2023Dec}%
  \BibitemOpen
  \bibfield  {author} {\bibinfo {author} {\bibfnamefont {M.}~\bibnamefont
  {Krug}}\ and\ \bibinfo {author} {\bibfnamefont {J.}~\bibnamefont
  {Stockburger}},\ }\bibfield  {title} {\bibinfo {title} {On stability issues
  of the heom method},\ }\href
  {https://doi.org/10.1140/epjs/s11734-023-00972-9} {\bibfield  {journal}
  {\bibinfo  {journal} {Eur. Phys. J. Spec. Top.}\ }\textbf {\bibinfo {volume}
  {232}},\ \bibinfo {pages} {3219} (\bibinfo {year} {2023})}\BibitemShut
  {NoStop}%
\bibitem [{\citenamefont {Alford}\ \emph {et~al.}(2025)\citenamefont {Alford},
  \citenamefont {Bettmann},\ and\ \citenamefont
  {Landi}}]{Alford_2_exponentials}%
  \BibitemOpen
  \bibfield  {author} {\bibinfo {author} {\bibfnamefont {W.}~\bibnamefont
  {Alford}}, \bibinfo {author} {\bibfnamefont {L.~P.}\ \bibnamefont
  {Bettmann}},\ and\ \bibinfo {author} {\bibfnamefont {G.~T.}\ \bibnamefont
  {Landi}},\ }\bibfield  {title} {\bibinfo {title} {Subtleties in the
  pseudomodes formalism},\ }\bibfield  {journal} {\bibinfo  {journal} {arXiv}\
  }\href {https://doi.org/10.48550/arXiv.2509.16377}
  {10.48550/arXiv.2509.16377} (\bibinfo {year} {2025}),\ \Eprint
  {https://arxiv.org/abs/2509.16377} {2509.16377} \BibitemShut {NoStop}%
\bibitem [{\citenamefont {Thoenniss}\ \emph {et~al.}(2025)\citenamefont
  {Thoenniss}, \citenamefont {Vilkoviskiy},\ and\ \citenamefont
  {Abanin}}]{quasi_lindblad_abanin_fermions}%
  \BibitemOpen
  \bibfield  {author} {\bibinfo {author} {\bibfnamefont {J.}~\bibnamefont
  {Thoenniss}}, \bibinfo {author} {\bibfnamefont {I.}~\bibnamefont
  {Vilkoviskiy}},\ and\ \bibinfo {author} {\bibfnamefont {D.~A.}\ \bibnamefont
  {Abanin}},\ }\bibfield  {title} {\bibinfo {title} {Efficient pseudomode
  representation and complexity of quantum impurity models},\ }\href
  {https://doi.org/10.1103/h8g7-bmng} {\bibfield  {journal} {\bibinfo
  {journal} {Phys. Rev. B}\ }\textbf {\bibinfo {volume} {112}},\ \bibinfo
  {pages} {155114} (\bibinfo {year} {2025})}\BibitemShut {NoStop}%
\bibitem [{\citenamefont {Yu}\ \emph {et~al.}(2026)\citenamefont {Yu},
  \citenamefont {Strunz},\ and\ \citenamefont {Nimmrichter}}]{Yu2026Jan}%
  \BibitemOpen
  \bibfield  {author} {\bibinfo {author} {\bibfnamefont {M.}~\bibnamefont
  {Yu}}, \bibinfo {author} {\bibfnamefont {W.~T.}\ \bibnamefont {Strunz}},\
  and\ \bibinfo {author} {\bibfnamefont {S.}~\bibnamefont {Nimmrichter}},\
  }\bibfield  {title} {\bibinfo {title} {Non-{M}arkovian dynamics of the giant
  atom beyond the rotating-wave approximation},\ }\bibfield  {journal}
  {\bibinfo  {journal} {arXiv}\ }\href
  {https://doi.org/10.48550/arXiv.2601.03383} {10.48550/arXiv.2601.03383}
  (\bibinfo {year} {2026}),\ \Eprint {https://arxiv.org/abs/2601.03383}
  {2601.03383} \BibitemShut {NoStop}%
\bibitem [{\citenamefont {Gao}\ \emph {et~al.}(2022)\citenamefont {Gao},
  \citenamefont {Ren}, \citenamefont {Eisfeld},\ and\ \citenamefont
  {Shuai}}]{dissiponsAlex}%
  \BibitemOpen
  \bibfield  {author} {\bibinfo {author} {\bibfnamefont {X.}~\bibnamefont
  {Gao}}, \bibinfo {author} {\bibfnamefont {J.}~\bibnamefont {Ren}}, \bibinfo
  {author} {\bibfnamefont {A.}~\bibnamefont {Eisfeld}},\ and\ \bibinfo {author}
  {\bibfnamefont {Z.}~\bibnamefont {Shuai}},\ }\bibfield  {title} {\bibinfo
  {title} {{Non-Markovian stochastic Schr\"odinger equation:
  Matrix-product-state approach to the hierarchy of pure states}},\ }\href
  {https://doi.org/10.1103/PhysRevA.105.L030202} {\bibfield  {journal}
  {\bibinfo  {journal} {Phys. Rev. A}\ }\textbf {\bibinfo {volume} {105}},\
  \bibinfo {pages} {L030202} (\bibinfo {year} {2022})}\BibitemShut {NoStop}%
\bibitem [{\citenamefont {Flannigan}\ \emph {et~al.}(2022)\citenamefont
  {Flannigan}, \citenamefont {Damanet},\ and\ \citenamefont
  {Daley}}]{dissiponsAndrew}%
  \BibitemOpen
  \bibfield  {author} {\bibinfo {author} {\bibfnamefont {S.}~\bibnamefont
  {Flannigan}}, \bibinfo {author} {\bibfnamefont {F.}~\bibnamefont {Damanet}},\
  and\ \bibinfo {author} {\bibfnamefont {A.~J.}\ \bibnamefont {Daley}},\
  }\bibfield  {title} {\bibinfo {title} {Many-body quantum state diffusion for
  non-{M}arkovian dynamics in strongly interacting systems},\ }\href
  {https://doi.org/10.1103/PhysRevLett.128.063601} {\bibfield  {journal}
  {\bibinfo  {journal} {Phys. Rev. Lett.}\ }\textbf {\bibinfo {volume} {128}},\
  \bibinfo {pages} {063601} (\bibinfo {year} {2022})}\BibitemShut {NoStop}%
\bibitem [{\citenamefont {Parthasarathy}(1992)}]{Parthasarathy1992}%
  \BibitemOpen
  \bibfield  {author} {\bibinfo {author} {\bibfnamefont {K.~R.}\ \bibnamefont
  {Parthasarathy}},\ }\href@noop {} {\emph {\bibinfo {title} {An Introduction
  to Quantum Stochastic Calculus}}}\ (\bibinfo  {publisher} {Springer},\
  \bibinfo {address} {Basel, Switzerland},\ \bibinfo {year} {1992})\BibitemShut
  {NoStop}%
\bibitem [{\citenamefont {Gardiner}\ and\ \citenamefont
  {Zoller}(2004)}]{Gardiner2004}%
  \BibitemOpen
  \bibfield  {author} {\bibinfo {author} {\bibfnamefont {C.}~\bibnamefont
  {Gardiner}}\ and\ \bibinfo {author} {\bibfnamefont {P.}~\bibnamefont
  {Zoller}},\ }\href {https://link.springer.com/book/9783540223016} {\emph
  {\bibinfo {title} {{Quantum Noise}}}}\ (\bibinfo  {publisher} {Springer},\
  \bibinfo {address} {Berlin, Germany},\ \bibinfo {year} {2004})\BibitemShut
  {NoStop}%
\bibitem [{\citenamefont {Müller}\ and\ \citenamefont
  {Strunz}()}]{supplement}%
  \BibitemOpen
  \bibfield  {author} {\bibinfo {author} {\bibfnamefont {K.}~\bibnamefont
  {Müller}}\ and\ \bibinfo {author} {\bibfnamefont {W.~T.}\ \bibnamefont
  {Strunz}},\ }\href@noop {} {\bibinfo {title} {Supplemental
  material}}\BibitemShut {NoStop}%
\bibitem [{\citenamefont {Sayed}\ and\ \citenamefont
  {Kailath}(2001)}]{Sayed2001Sep}%
  \BibitemOpen
  \bibfield  {author} {\bibinfo {author} {\bibfnamefont {A.~H.}\ \bibnamefont
  {Sayed}}\ and\ \bibinfo {author} {\bibfnamefont {T.}~\bibnamefont
  {Kailath}},\ }\bibfield  {title} {\bibinfo {title} {A survey of spectral
  factorization methods},\ }\href {https://doi.org/10.1002/nla.250} {\bibfield
  {journal} {\bibinfo  {journal} {Numer. Linear Algebra Appl.}\ }\textbf
  {\bibinfo {volume} {8}},\ \bibinfo {pages} {467} (\bibinfo {year}
  {2001})}\BibitemShut {NoStop}%
\bibitem [{\citenamefont {Mahoney}\ and\ \citenamefont
  {Sivazlian}(1983)}]{Mahoney1983Sep}%
  \BibitemOpen
  \bibfield  {author} {\bibinfo {author} {\bibfnamefont {J.~F.}\ \bibnamefont
  {Mahoney}}\ and\ \bibinfo {author} {\bibfnamefont {B.~D.}\ \bibnamefont
  {Sivazlian}},\ }\bibfield  {title} {\bibinfo {title} {Partial fractions
  expansion: a review of computational methodology and efficiency},\ }\href
  {https://doi.org/10.1016/0377-0427(83)90018-3} {\bibfield  {journal}
  {\bibinfo  {journal} {J. Comput. Appl. Math.}\ }\textbf {\bibinfo {volume}
  {9}},\ \bibinfo {pages} {247} (\bibinfo {year} {1983})}\BibitemShut {NoStop}%
\bibitem [{\citenamefont {Park}\ \emph {et~al.}(2024)\citenamefont {Park},
  \citenamefont {Huang}, \citenamefont {Zhu}, \citenamefont {Yang},
  \citenamefont {Chan},\ and\ \citenamefont {Lin}}]{quasi_lindblad_Lin}%
  \BibitemOpen
  \bibfield  {author} {\bibinfo {author} {\bibfnamefont {G.}~\bibnamefont
  {Park}}, \bibinfo {author} {\bibfnamefont {Z.}~\bibnamefont {Huang}},
  \bibinfo {author} {\bibfnamefont {Y.}~\bibnamefont {Zhu}}, \bibinfo {author}
  {\bibfnamefont {C.}~\bibnamefont {Yang}}, \bibinfo {author} {\bibfnamefont
  {G.~K.-L.}\ \bibnamefont {Chan}},\ and\ \bibinfo {author} {\bibfnamefont
  {L.}~\bibnamefont {Lin}},\ }\bibfield  {title} {\bibinfo {title}
  {Quasi-lindblad pseudomode theory for open quantum systems},\ }\href
  {https://doi.org/10.1103/PhysRevB.110.195148} {\bibfield  {journal} {\bibinfo
   {journal} {Phys. Rev. B}\ }\textbf {\bibinfo {volume} {110}},\ \bibinfo
  {pages} {195148} (\bibinfo {year} {2024})}\BibitemShut {NoStop}%
\bibitem [{\citenamefont {Wiseman}\ and\ \citenamefont
  {Milburn}(2010)}]{Wiseman2010}%
  \BibitemOpen
  \bibfield  {author} {\bibinfo {author} {\bibfnamefont {H.~M.}\ \bibnamefont
  {Wiseman}}\ and\ \bibinfo {author} {\bibfnamefont {G.~J.}\ \bibnamefont
  {Milburn}},\ }\href@noop {} {\emph {\bibinfo {title} {Quantum Measurement and
  Control}}}\ (\bibinfo  {publisher} {Cambridge University Press},\ \bibinfo
  {address} {Cambridge, England, UK},\ \bibinfo {year} {2010})\BibitemShut
  {NoStop}%
\bibitem [{\citenamefont {Dalibard}\ \emph {et~al.}(1992)\citenamefont
  {Dalibard}, \citenamefont {Castin},\ and\ \citenamefont
  {M{\o}lmer}}]{Dalibard1992Feb}%
  \BibitemOpen
  \bibfield  {author} {\bibinfo {author} {\bibfnamefont {J.}~\bibnamefont
  {Dalibard}}, \bibinfo {author} {\bibfnamefont {Y.}~\bibnamefont {Castin}},\
  and\ \bibinfo {author} {\bibfnamefont {K.}~\bibnamefont {M{\o}lmer}},\
  }\bibfield  {title} {\bibinfo {title} {Wave-function approach to dissipative
  processes in quantum optics},\ }\href
  {https://doi.org/10.1103/PhysRevLett.68.580} {\bibfield  {journal} {\bibinfo
  {journal} {Phys. Rev. Lett.}\ }\textbf {\bibinfo {volume} {68}},\ \bibinfo
  {pages} {580} (\bibinfo {year} {1992})}\BibitemShut {NoStop}%
\bibitem [{\citenamefont {Carmichael}(1993)}]{Carmichael1993}%
  \BibitemOpen
  \bibfield  {author} {\bibinfo {author} {\bibfnamefont {H.}~\bibnamefont
  {Carmichael}},\ }\href {https://doi.org/10.1007/978-3-540-47620-7} {\emph
  {\bibinfo {title} {An Open Systems Approach to Quantum Optics}}}\ (\bibinfo
  {publisher} {Springer},\ \bibinfo {address} {Berlin, Germany},\ \bibinfo
  {year} {1993})\BibitemShut {NoStop}%
\bibitem [{\citenamefont {Sukharnikov}\ \emph {et~al.}(2026)\citenamefont
  {Sukharnikov}, \citenamefont {Chuchurka},\ and\ \citenamefont
  {Schlawin}}]{Sukharnikov2026Feb}%
  \BibitemOpen
  \bibfield  {author} {\bibinfo {author} {\bibfnamefont {V.}~\bibnamefont
  {Sukharnikov}}, \bibinfo {author} {\bibfnamefont {S.}~\bibnamefont
  {Chuchurka}},\ and\ \bibinfo {author} {\bibfnamefont {F.}~\bibnamefont
  {Schlawin}},\ }\bibfield  {title} {\bibinfo {title} {{Non-Markovian dynamics
  in nonstationary Gaussian baths: A hierarchy of pure states approach}},\
  }\href {https://doi.org/10.1103/yt37-s9hz} {\bibfield  {journal} {\bibinfo
  {journal} {Phys. Rev. Res.}\ }\textbf {\bibinfo {volume} {8}},\ \bibinfo
  {pages} {013123} (\bibinfo {year} {2026})}\BibitemShut {NoStop}%
\bibitem [{\citenamefont {Parks}(1992)}]{Parks1992Jan}%
  \BibitemOpen
  \bibfield  {author} {\bibinfo {author} {\bibfnamefont {P.~C.}\ \bibnamefont
  {Parks}},\ }\bibfield  {title} {\bibinfo {title} {{A. M. Lyapunov's}
  stability theory{\ifmmode---\else\textemdash\fi}100 years on {$\ast$}},\
  }\href {https://doi.org/10.1093/imamci/9.4.275} {\bibfield  {journal}
  {\bibinfo  {journal} {IMA J. Math. Control Inf.}\ }\textbf {\bibinfo {volume}
  {9}},\ \bibinfo {pages} {275} (\bibinfo {year} {1992})}\BibitemShut {NoStop}%
\bibitem [{\citenamefont {Heuser}(2006)}]{heuserAnalysis}%
  \BibitemOpen
  \bibfield  {author} {\bibinfo {author} {\bibfnamefont {H.}~\bibnamefont
  {Heuser}},\ }\href@noop {} {\emph {\bibinfo {title} {{Lehrbuch der Analysis}
  1}}},\ \bibinfo {edition} {16th}\ ed.\ (\bibinfo  {publisher} {Teubner},\
  \bibinfo {address} {Wiesbaden},\ \bibinfo {year} {2006})\BibitemShut
  {NoStop}%
\bibitem [{\citenamefont
  {Schollw{\ifmmode\ddot{o}\else\"{o}\fi}ck}(2011)}]{Schollwock2011Jan}%
  \BibitemOpen
  \bibfield  {author} {\bibinfo {author} {\bibfnamefont {U.}~\bibnamefont
  {Schollw{\ifmmode\ddot{o}\else\"{o}\fi}ck}},\ }\bibfield  {title} {\bibinfo
  {title} {{The density-matrix renormalization group in the age of matrix
  product states}},\ }\href {https://doi.org/10.1016/j.aop.2010.09.012}
  {\bibfield  {journal} {\bibinfo  {journal} {Ann. Phys.}\ }\textbf {\bibinfo
  {volume} {326}},\ \bibinfo {pages} {96} (\bibinfo {year} {2011})}\BibitemShut
  {NoStop}%
\bibitem [{\citenamefont {Di\'osi}\ \emph {et~al.}(1998)\citenamefont
  {Di\'osi}, \citenamefont {Gisin},\ and\ \citenamefont {Strunz}}]{NMQSD}%
  \BibitemOpen
  \bibfield  {author} {\bibinfo {author} {\bibfnamefont {L.}~\bibnamefont
  {Di\'osi}}, \bibinfo {author} {\bibfnamefont {N.}~\bibnamefont {Gisin}},\
  and\ \bibinfo {author} {\bibfnamefont {W.~T.}\ \bibnamefont {Strunz}},\
  }\bibfield  {title} {\bibinfo {title} {Non-markovian quantum state
  diffusion},\ }\href {https://doi.org/10.1103/PhysRevA.58.1699} {\bibfield
  {journal} {\bibinfo  {journal} {Phys. Rev. A}\ }\textbf {\bibinfo {volume}
  {58}},\ \bibinfo {pages} {1699} (\bibinfo {year} {1998})}\BibitemShut
  {NoStop}%
\bibitem [{\citenamefont {Hartmann}(2021)}]{richardThesis}%
  \BibitemOpen
  \bibfield  {author} {\bibinfo {author} {\bibfnamefont {R.}~\bibnamefont
  {Hartmann}},\ }\emph {\bibinfo {title} {Exact Open Quantum System Dynamics -
  Investigating Environmentally Induced Entanglement}},\ \href@noop {} {Ph.D.
  thesis},\ \bibinfo  {school} {Technische Universität Dresden} (\bibinfo
  {year} {2021})\BibitemShut {NoStop}%
\bibitem [{\citenamefont {Boettcher}\ \emph {et~al.}(2024)\citenamefont
  {Boettcher}, \citenamefont {Hartmann}, \citenamefont {Beyer},\ and\
  \citenamefont {Strunz}}]{Boettcher2024Mar}%
  \BibitemOpen
  \bibfield  {author} {\bibinfo {author} {\bibfnamefont {V.}~\bibnamefont
  {Boettcher}}, \bibinfo {author} {\bibfnamefont {R.}~\bibnamefont {Hartmann}},
  \bibinfo {author} {\bibfnamefont {K.}~\bibnamefont {Beyer}},\ and\ \bibinfo
  {author} {\bibfnamefont {W.~T.}\ \bibnamefont {Strunz}},\ }\bibfield  {title}
  {\bibinfo {title} {Dynamics of a strongly coupled quantum heat
  engine{\ifmmode---\else\textemdash\fi}computing bath observables from the
  hierarchy of pure states},\ }\bibfield  {journal} {\bibinfo  {journal} {J.
  Chem. Phys.}\ }\textbf {\bibinfo {volume} {160}},\ \href
  {https://doi.org/10.1063/5.0192075} {10.1063/5.0192075} (\bibinfo {year}
  {2024})\BibitemShut {NoStop}%
\end{thebibliography}%

\clearpage
\onecolumngrid

\setcounter{equation}{0}
\setcounter{section}{0}
\renewcommand{\theequation}{S\arabic{equation}}
\renewcommand{\thesection}{\arabic{section}}
\renewcommand{\thefigure}{S\arabic{figure}}
\renewcommand{\i}{\mathrm{i}}

\newcommand{\et}[1]{|#1)}
\newcommand{\ra}[1]{(#1|}
\newcommand{\raet}[2]{(#1|#2)}
\newcommand{\ev}[1]{\langle\!\langle\;#1\;\rangle\!\rangle}
\newcommand{\opone}{1\!\!\!1}

\setcounter{secnumdepth}{2}

\begin{center}
{\large \bf Supplemental Material}
\end{center}

In the following we present complete proofs of the results stated in the main text. After discussing the interaction picture of the pseudomode Hamiltonian, we give an overview over multivariate Ornstein-Uhlenbeck processes. \Kadd{Following these preliminaries, we then show in Sec.~\ref{sec:representability}} that every physical exponential bath correlation function (BCF) can be realized by a pseudomode model (pseudomode representability).
Finally, \Kadd{in Sec.~\ref{sec:dissipon}}, we introduce the dissipon transformation, which maps the system and pseudomode state onto the HEOM and HOPS states, establishing the one-to-one correspondence between the two approaches.

\section{Pseudomode interaction picture}
As mentioned in the main text the pseudomode method uses an Ansatz-Lindbladian describing damped, coupled harmonic oscillators to approximate the bath correlation function (BCF) of the original Gaussian environment.
The system and pseudomode dynamics is given by the Lindbladian in Eq.~(6) as 
\begin{equation}\label{eq:SIdefPseudomode}
    \begin{split}
    \dot{\rho} =& -i[H_{\mathrm{pm}},\rho] + \sum_k L_k\rho L_k^\dagger -\frac{1}{2}\{L_k^\dagger L_k, \rho\},\\
        H_{\mathrm{pm}} =& H_{\mathrm{sys}} + S\otimes(A + A^\dagger)  + \sum_{k,k'}h_{k,k'}a_k^\dagger a_{k'},
    \end{split}
\end{equation}
with $A = \sum_k g_k^* a_k$ and the Lindbladians $L_k = \sum_{k'} \Gamma_{k,k'} a_{k'}$.
For our proofs it turns out useful to switch from a reduced description in terms of the Lindbladian \eqref{eq:SIdefPseudomode} to the equivalent formalism of quantum stochastic calculus \cite{Parthasarathy1992, Gardiner2004}. There, we describe the unitary evolution in the total Hilbert space, which explicitly includes the Markovian baths of each pseudomode in terms of an operator quantum white noise.
In this total Hilbert space the pseudomode ansatz can be brought into the form of Eq.~(1) in the main text, as we will show in the following.
The unitary evolution in the total Hilbert space of system, pseudomodes and Markovian baths is determined by 
\begin{equation}\label{eq:fundamentalHCQED}
    \Htot = H_{\mathrm{pm}} + \sum_k \left(L_k \xi_k^\dagger + L_k^\dagger \xi_k\right) + \sum_{k,\lambda}  \omega_{k,\lambda}\, d_{k,\lambda}^\dagger d_{k,\lambda},
\end{equation}
where $\xi_k = \sum_\lambda \eta_{k,\lambda}^* d_{k,\lambda}$ is a Markov environment coupling agent, linear in the Markov environment annihilation operators $d_k$. 
Crucially, in the usual Markov limit, \Kadd{the parameters $\eta_{k,\lambda},\,\omega_{k,\lambda}$ satisfy $\sum_\lambda |\eta_{k,\lambda}|^2e^{-i\omega_{k,\lambda}\tau} = \delta(\tau)$, such that the} (non-Hermitian) coupling operator in Heisenberg picture, $\xi_k(t) = \sum_\lambda \eta_{k,\lambda}^* d_{k,\lambda} e^{-\i \omega_{k,\lambda} t}$, 
\Kadd{becomes} operator white noise, with vacuum correlation function (or commutation relation) \cite{Parthasarathy1992,Gardiner2004}
\begin{equation}\label{eq:SIoperatorwhitenoise}
\langle \xi_k(t)\xi_{k'}^\dagger(s)\rangle_{\mathrm{vac}} = [\xi_k(t),\xi_{k'}^\dagger(s)]= \delta_{k,k'}\delta(t-s).
\end{equation}
The full environment of the pseudomode model includes the modes $a_k$ and $d_k$ and we thus define
\begin{equation}\label{eq:HenvCQED}
    \Henv = \sum_{k,k'}h_{k,k'} a_k^\dagger a_{k'} + \sum_k \left(L_k \xi_k^\dagger + L_k^\dagger \xi_k\right) + \sum_{k,\lambda}  \omega_{k,\lambda}\, d_{k,\lambda}^\dagger d_{k,\lambda}.
\end{equation}
In an interaction picture with respect to this $\Henv$ the Hamiltonian \eqref{eq:fundamentalHCQED} with $H_{\mathrm{pm}}$ from \eqref{eq:SIdefPseudomode} thus takes the form of Eq.~(1) in the main text, with $B(t) = A(t)$, where $A(t) = \sum_k g_k^* a_k(t)$ follows from the Heisenberg equations of motion for the pseudomodes $\i \dot a_k = [a_k,\Henv]$ with $\Henv$ from (\ref{eq:HenvCQED}). 
Combined with the corresponding Heisenberg equation of motion for the modes $d_{k,\lambda}$, this differential equation leads to an operator Ornstein-Uhlenbeck process with operator white noise from (\ref{eq:SIoperatorwhitenoise}). Thus, the interaction picture dynamics given in Eq.~(7) ensues,
\begin{equation}\label{eq:SIpseudomodeOU}
    \begin{split}
        H_{\mathrm{pm,tot}} =& H_{\mathrm{sys}} + S\otimes(\hat A(t)+\hat A^\dagger(t)),\;\;\;\;\;A(t) = \sum_k g_k^* a_k(t)\\
        \dot{a}_k(t) =& \sum_{k'}-(ih_{k,k'} +\frac{1}{2}(\Gamma^\dagger\Gamma)_{k,k'})a_{k'} + \Gamma^\dagger_{k,k'}{\xi}_{k'}(t).
    \end{split}
\end{equation}
\Kadd{Our ultimate goal in Sec.~\ref{sec:OUprocesses}-\ref{sec:representability} is to find parameters $h,\,\Gamma,\,\bvec g$ that reproduce a given exponential BCF}
\begin{equation}\label{eq:Abcf}
    \langle{\hat A(t) \hat A^\dagger(s)}\rangle = \alpha_{exp}(\tau).
\end{equation}

\section{General results on multivariate Ornstein-Uhlenbeck processes}\label{sec:OUprocesses}

In the following we present an analysis of (c-number) multivariate Ornstein-Uhlenbeck (OU) processes $\bvec{z}(t) := (z_1(t),z_2(t),\ldots,z_N(t))^T \in \mathbb{C}^N$ for an $N$-dimensional, complex OU-vector. 
\Kadd{An insertion of the identity in terms of coherent states in Eq.~\eqref{eq:Abcf} makes it clear that the pseudomode BCF is identical to the correlation function of the c-numbers process, if the $\bvec z(t)$ fulfill the same differential equation. Thus, if we succeed in finding a c-number process $\bvec z(t)$ we have also found a pseudomode model.}

\subsection{Differential equation, solution, and correlation function}

We start with the generic stochastic differential equation of an Ornstein-Uhlenbeck process,
\begin{equation}\label{eq:oupro}
    {\bvec{\dot z}}(t) = -M\bvec{z}(t) + B \bvec{\xi}(t),
\end{equation}
where $M$ is a matrix that leads to a stable solution (i.e. $||\exp{(-Mt)}||\rightarrow 0$ for $t\rightarrow \infty$), and the matrix $B$ determines the diffusion matrix. Here, $\bvec{\xi}(t):= (\xi_1(t),\xi_2(t),\ldots,\xi_N(t))^T$ is a vector of complex, independent white noises, i.e.
\begin{equation}
    \ev{\xi_n(t) \xi_m^*(s)} = \delta_{nm}\delta(t-s), \;\;\;  \ev{\xi_n(t) \xi_m(s)} = 0,\;\;\;  \ev{\xi_n(t)} = 0.
\end{equation}
Most conveniently, we can write instead
\begin{equation}
    \ev{\bvec{\xi}(t)\bvec{\xi}(s)^\dagger} = \opone\delta(t-s).
\end{equation}
The solution of the OU-differential equation is given by
\begin{equation}\label{eq:sol}
    \bvec{z}(t) = e^{-Mt} \bvec{z}(0) + \int_0^t\!ds\, e^{-M(t-s)}\,B\,\bvec{\xi}(s).
\end{equation}
In the following, w.l.o.g., we always assume $t>s$.
From the explicit solution \eqref{eq:sol}, it is straightforward to obtain the correlation expression
\begin{equation}\label{eq:cor1}
     \ev{\bvec{z}(t)\bvec{z}^\dagger(s)} = e^{-Mt}\ev{\bvec{z}(0)\bvec{z}^\dagger(0)} e^{-M^\dagger s} + e^{-M(t-s)}\int_0^s\!ds'\, e^{-Ms'}\,BB^\dagger  e^{-M^\dagger s'}.
\end{equation}
Introducing the asymptotic, obviously positive and Hermitian matrix
\begin{equation}\label{eq:pmat}
    P = \int_0^\infty\!ds'\, e^{-Ms'}\,BB^\dagger  e^{-M^\dagger s'},
\end{equation}
the correlation matrix \eqref{eq:cor1} takes the more appealing form
\begin{equation}\label{eq:cor2}
 \ev{\bvec{z}(t)\bvec{z}^\dagger(s)} =
      e^{-M(t-s)}P + e^{-Mt}\left(\ev{\bvec{z}(0)\bvec{z}^\dagger(0)}-P\right)e^{-M^\dagger s}.
\end{equation}
With the right choice of random initial conditions, $\ev{\bvec{z}(0)\bvec{z}^\dagger(0)}=P$, (alternatively, waiting for the stationary regime), we obtain the stationary correlation function
\begin{equation}\label{eq:cors}
     \ev{\bvec{z}(t)\bvec{z}^\dagger(s)} = e^{-M(t-s)}P\;.
\end{equation}
Note that from \eqref{eq:pmat} we find
\begin{align}\label{eq:lyap}
    MP+PM^\dagger & = 
    - \int_0^\infty\!ds'\,\left(\frac{d}{ds'}\,\left(e^{-Ms'}\,BB^\dagger  e^{-M^\dagger s'}\right)\right) 
    = BB^\dagger,
\end{align}
a result reminiscent of Lyapunov's \Kadd{matrix equation} \cite{Parks1992Jan}, which will become relevant shortly.
Ultimately, with the combined process $A(t) = \sum_k \Tilde g_k^* z_k(t) = \bvec{\Tilde g}^\dagger\bvec{z}(t)$, we are interested in the correlation function
\begin{equation}\label{eq:corf}
    \alpha(t-s) = \ev{A(t)A^*(s)} = 
    \ev{\bvec{\Tilde g}^\dagger \bvec{z}(t)
    \bvec{z}^\dagger(s)\bvec{\Tilde g}} = \bvec{\Tilde g}^\dagger e^{-M(t-s)}P\bvec{\Tilde g}.
\end{equation}

\subsection{Ornstein-Uhlenbeck process from a pseudomode model}
A pseudomode model leads to a particular OU-process \eqref{eq:SIpseudomodeOU}, where the matrices defining the OU process in \eqref{eq:oupro} are given by
\begin{align}\label{eq:pseudocond}
    \tilde{M} = \frac{1}{2}\Gamma^\dagger\Gamma + \i H,\;\;\;\; \tilde{B}=\Gamma^\dagger,
\end{align}
so that the two matrices cannot be chosen independently. 
Interestingly, 
in this case it turns out that the matrix $\tilde{P}$ in \eqref{eq:pmat} is trivial, ${\tilde{P}}=\opone$. This can be seen as follows: using $\tilde B\tilde B^\dagger= \tilde M+{\tilde M}^\dagger$, we find
\begin{align}
    \tilde{P} = \int_0^\infty\!ds'\, e^{-\tilde Ms'}\,\tilde B\tilde B^\dagger  e^{-\tilde M^\dagger s'} 
    = - \int_0^\infty\!ds'\, \left(\frac{d}{ds'} e^{-\tilde Ms'}\, e^{-\tilde M^\dagger s'}\right)
    = \opone.
\end{align}

Thus, instead of \eqref{eq:cors}, the pseudomode-model correlation function simplifies to
\begin{align}\label{eq:corspm}
    \ev{\bvec{\tilde z}(t)\bvec{\tilde z}^\dagger(s)}=e^{-\tilde M(t-s)},
\end{align}
and the bath correlation function for $A(t) = \sum_k g_k^* \tilde z_k(t) = \bvec{g}^\dagger\bvec{\tilde z}(t)$ takes the elegant form (see main text and \cite{pseudomodes_lin})
\begin{equation}\label{eq:pseudomodeBCFelegant}
    \alpha(t-s) =  \bvec{g}^\dagger e^{-\tilde M(t-s)}\bvec{ g}.
\end{equation}

\subsection{Pseudomode transformation}\label{ssec:pseudomodeTrafo}
We show here that with a linear transformation $V$, \emph{any} Ornstein-Uhlenbeck process (matrices $M$ and $B$ in \eqref{eq:oupro}) can be transformed into an OU process emerging from a pseudomode model with matrices $\tilde M$ and $\tilde B$ as in \eqref{eq:pseudocond}. We start with
a linear (invertible) matrix $V$ and define
\begin{align}
\bvec{\tilde z}(t) = V\bvec{z}(t),
\end{align}
which allows to transform the matrices in the OU differential equation. From \eqref{eq:oupro} we find for $\bvec{\tilde z}(t)$
\begin{equation}\label{eq:noupro}
    \frac{d}{dt}\bvec{\tilde z}(t) = -VMV^{-1}\bvec{\tilde z}(t) + VB \bvec{\xi}(t) = -\tilde M\bvec{\tilde z}(t) + \tilde B \bvec{\xi}(t),
\end{equation}
so that in order to satisfy \eqref{eq:pseudocond} we search a $V$ with
\begin{align}\label{eq:vcond}
    VMV^{-1} = \frac{1}{2}\Gamma^\dagger\Gamma + \i h,\;\;\;\;VB = \Gamma^\dagger.
\end{align}
Inserting the second condition into the first, and eliminating the Hamiltonian by adding the adjoint of that equation, results in the condition for $V$,
\begin{align}
    M(V^\dagger V)^{-1} + (V^\dagger V)^{-1} M^\dagger = BB^\dagger.
\end{align}
A glance at \eqref{eq:lyap} reveals that we can satisfy this relation with the choice
\begin{equation}\label{eq:vandp}
    (V^\dagger V)^{-1} = P,
\end{equation}
where $P$ from \eqref{eq:pmat} was a positive, Hermitian operator and thus such a $V$ always exists. In fact, writing
\begin{equation}\label{eq:diagp}
    P = W\mathcal{D}W^\dagger,
\end{equation}
where $W$ is unitary and $\mathcal{D}$ diagonal with the (positive, real) eigenvalues of $P$ along its diagonal, we can chose
\begin{equation}\label{eq:defv}
    V = U\mathcal{D}^{-1/2} W^\dagger
\end{equation}
with an arbitrary, unitary $U$.

Having found the pseudomode model OU process $\bvec{\tilde z}(t)$, we need to make sure that we arrive at the correct correlation function $\alpha(t-s)$. We can simply write from \eqref{eq:corf}
\begin{equation}
    \alpha(t-s) = \ev{\bvec{\Tilde g}^\dagger\bvec{z}(t)
    \bvec{z}^\dagger(s)\bvec{\Tilde g}} = \ev{\bvec{\Tilde g}^\dagger V^{-1} V\bvec{z}(t)
    \bvec{z}^\dagger(s)V^\dagger(V^{-1})^\dagger \bvec{\Tilde g}} = \ev{\bvec{g}^\dagger\bvec{\tilde z}(t)\bvec{\tilde z}^\dagger(s)\bvec{g}},
\end{equation}
so that the desired correlation function can be obtained from the pseudomode OU-process \eqref{eq:noupro} by a linear transformation of the coupling constants,
\begin{equation}\label{eq:gTrafo}
    \bvec{g} = (V^{-1})^\dagger\bvec{\Tilde g}.
\end{equation}

\Kadd{In summary, the results of this section have simplified our task to finding \emph{any} Ornstein-Uhlenbeck process (arbitrary $M,\,B$) with correlation $\alpha(\tau) = \alpha_{exp}(\tau)$. The pseudomode parameters $h,\,\Gamma,\,\bvec g$ can then be obtain via the above transformation.}

\section{Proof of pseudomode representability}\label{sec:representability}
We start from a bath correlation function (BCF) that can be expressed as a sum of exponential terms
\begin{equation}\label{eq:exponentialcorrelation2}
    \alpha_{\mathrm{exp}}(\tau) = \sum_{j=1}^N G_je^{-\lambda_j\tau},\quad G_j,\,\lambda_j \in \mathbb{C},\;\;\;\;\mbox{for}\;\;\tau\ge 0,
\end{equation}
with $\lambda_j=\gamma_j+\i\omega_j$ and positive real parts $\gamma_j>0$, and we set 
$\alpha_{\mathrm{exp}}(\tau) = \alpha_{\mathrm{exp}}(-\tau)^*$ for $\tau\le 0$, as required from the fundamental relation Eq.~(2). Note that this implies
$\alpha_{\mathrm{exp}}(0)=\alpha_{\mathrm{exp}}(0)^*$ or 
\begin{equation}\label{eq:realcondition}
  \sum_j\operatorname{Im}(G_j) = 0.  
\end{equation}

As the BCF of a single pseudomode is given by an exponential with a real positive amplitude $G_j$, it has always been known that every BCF expressed as a sum of $N$ exponential terms with \emph{real and positive} prefactors $G_j>0$ can be obtained from $N$ \emph{uncoupled} pseudomodes.
However, by allowing interactions between pseudomodes it is also possible to obtain negative or complex $G_j$, the "band gap model" of \cite{Garraway1997Mar} being an example. In fact, using general, complex $G_j$ in the Ansatz \eqref{eq:exponentialcorrelation2} significantly improves the efficiency of the bath correlation fit, from $\mathrm{poly}(T/\epsilon)$ to $\mathrm{polylog}(T/\epsilon)$ in the error $\epsilon$ and simulation time $T$ \cite{pseudomodes_lin, quasi_lindblad_abanin_fermions}.

We answer the following question with the proof below: can every \emph{physical} correlation function (i.e. those representing positive kernels) given by a sum of $N$ exponentials with \emph{complex} prefactors $G_j$ as in \eqref{eq:exponentialcorrelation2}, be obtained from a collection of $N$ coupled pseudomodes, as in Eq.~\eqref{eq:SIdefPseudomode}? The answer is affirmative.\\
We will proceed in two steps: First, we show the conditions that arise from the positive kernel condition on the parameters in Eq.~\eqref{eq:exponentialcorrelation2}. 
Second, we prove that these conditions are sufficient to guarantee the existence of an 
OU process that reproduces the given BCF, which can then be brought into the form of a pseudomode model Eq.~\eqref{eq:SIpseudomodeOU} according to Sec.~\ref{ssec:pseudomodeTrafo}.

\subsection{Positive kernel condition}\label{ssec:kernelCondition}
As discussed in the main text, a BCF is physical if it corresponds to a non-negative spectral density. 
For BCFs given as a sum of exponentials \eqref{eq:exponentialcorrelation2} the spectral density reads
\begin{align}
    J(\omega) =& \int_{-\infty}^{\infty}\alpha_{\mathrm{exp}}(\tau)e^{i\omega\tau}\dd \tau,\;\;
    = \sum_{j=1}^{N} \left(\frac{G_j}{\lambda_j -i\omega} \;+\; \mbox{c.c.}\right),\label{eq:J_form}\notag \\
    =& 2\; \frac{\sum_j\Big(\operatorname{Re}(G_j\lambda_j^*) - \omega\operatorname{Im}(G_j)\Big)\prod_{j'\neq j}(\lambda_{j'} - i\omega)(\lambda_{j'}^*+i\omega)}{\prod_j (\lambda_j - i\omega)(\lambda_j^*+i\omega)},\notag\\
    \equiv& \frac{P(\omega)}{Q(\omega)} \geq 0\quad\forall\omega\in \mathbb{R},\notag
\end{align}
where we assume w.l.o.g. that all $\lambda_j$ are distinct. 
We have expressed $J(\omega)$ as a rational function, where $Q(\omega)$ is a polynomial of degree $2N$ and $P(\omega)$ is a polynomial of degree $2(N-1)$. The term of order $\omega^{2N-1}$ in $P(\omega)$ is proportional to $\sum_j\operatorname{Im}(G_j)$ and this vanishes due to \eqref{eq:realcondition}. 
Since clearly $Q(\omega)\geq0$, the positive kernel condition requires $P(\omega)\geq0$, which implies that all real zeros $\Omega^R_\ell$ of $P$ must come at an even multiplicity $2m_\ell$. Additionally, it follows from $P(\omega) = P(\omega)^*$ that all complex zeros must come in complex conjugate pairs.
Thus, we can sort the complex zeros into two groups of equal size, where one groups contains all zeros in the upper half of the complex plane $\Omega^C_n$ and the second group the symmetric zeros in the lower half $\Omega^{C*}_n$.
We can then write
\begin{equation*}
    \begin{split}
        J(\omega) =& v(\omega)v^*(\omega),\\
        v(\omega) =& c\cdot\frac{\prod_\ell(\omega-\Omega^R_\ell)^{m_\ell}\prod_n(\omega-\Omega_n^C)}{\prod_j(\lambda_j - i\omega)},
    \end{split}
\end{equation*}
with some $c\in\mathbb{C}$.
Therefore, $v(\omega)$ is a rational function, where the degree of the numerator ($N-1$) is lower than the degree of the denominator ($N$). 
Due to the partial fraction decomposition theorem \cite{heuserAnalysis} we are thus guaranteed that it can be expanded as
\begin{equation}\label{eq:partialfrac}
    v(\omega) = \sum_{j=1}^N\frac{r_j}{\lambda_j -i\omega},\quad r_j\in\mathbb{C}.
\end{equation}
In summary, we have demonstrated that the positive kernel condition necessitates $J(\omega)$ being of the following form:
\begin{equation}\label{eq:JinMForm}
    J(\omega) = 
    \sum_{j,k=1}^{N}\frac{r_jr_k^*}{(\lambda_j -i\omega)(\lambda_k^* +i\omega)}.
\end{equation}
Now we write
\begin{equation}
    \frac{1}{(\lambda_j -i\omega)}\frac{1}{(\lambda_k^* +i\omega)}= \frac{1}{\lambda_j + \lambda_k^*}\left(\frac{1}{\lambda_j -i\omega}+ \frac{1}{\lambda_k^* +i\omega}\right),
\end{equation}
and evaluate
\begin{equation}
    \alpha(\tau) = \frac{1}{2\pi}\int_{-\infty}^\infty J(\omega) e^{-i\omega\tau}\,\dd \omega 
\end{equation} 
with the residue theorem -- closing the contour either in the upper or lower half plane depending on the sign of $\tau$. In this way we recover expression \eqref{eq:exponentialcorrelation2} with the complex amplitudes $G_j$ expressed in terms of the coefficients $r_j$ of the partial fraction decomposition \eqref{eq:partialfrac},
\begin{equation}
    G_j = \sum_k\frac{r_j r_k^*}{\lambda_j+\lambda_k^*},
\end{equation}
as claimed in the main text. Thus, from \eqref{eq:exponentialcorrelation2}, the bath correlation function (being a positive kernel) can always be written in 
the form
\begin{equation}\label{eq:corf2}
    \alpha(t-s) = \sum_{j,k=1}^N \frac{r_j r_k^*}{\lambda_j+\lambda_k^*}e^{-\lambda_j(t-s)}.
\end{equation}

\subsection{Corresponding Ornstein-Uhlenbeck process}

With expression \eqref{eq:corf2} for $\alpha(t-s)$ at hand, we can easily construct a corresponding Ornstein-Uhlenbeck process. It turns out sufficient to start with the following special, simple type of multivariate process
\begin{equation}\label{eq:oudiag}
    \dot z_j(t) = -\lambda_j z_j(t) + c_j \xi(t),\;\;\;j=1,\ldots,N\, ,
\end{equation}
where $\operatorname{Re}(\lambda_j)\ge 0$, and $\xi(t)$ is a \emph{single}, complex white noise. Clearly, this process is of the type \eqref{eq:oupro}, with a diagonal matrix $M=$diag$(\lambda_j)$ and $B=\bvec{c}\bvec{\phi}^\dagger$, where $\bvec{\phi}$  is a normalized vector such that $\xi(t)=\bvec{\phi}^\dagger \bvec{\xi(t)}$ is a single white noise process. From \eqref{eq:corf}, the correlation function of $A(t) = \bvec{\Tilde g}^\dagger\bvec{z}(t)$ is $\alpha(t-s) = \bvec{\Tilde g}^\dagger e^{-M(t-s)}P\bvec{\Tilde g}$. Working in the basis in which $M$ is diagonal, we find from \eqref{eq:pmat} $P_{jk} = \frac{c_j c_k^*}{\lambda_j + \lambda_k^*}$ and thus  from \eqref{eq:corf} the expression
\begin{equation}
\alpha(t-s) = \ev{A(t) A^*(s)} = \sum_{jk}
\frac{\Tilde g_j^*c_j c_k^*\Tilde g_k}{\lambda_j + \lambda_k^*}e^{-\lambda_j(t-s)}.
\end{equation}
We see that the desired bath correlation function \eqref{eq:corf2} can be obtained from the process \eqref{eq:oudiag} with vector $\bvec{\Tilde g}$ in multiple ways, as long as the relation
\begin{equation}\label{eq:gTilde_rRelation}
    \Tilde g_j^* c_j = r_j\;\;\mbox{for}\;\;j=1,\ldots,N
\end{equation}
is satisfied. 
We use this freedom \Kadd{here in a way that allows us to directly obtain the HEOM Eq.~(5) (with balanced "couplings" $\sqrt{G_j}$) later in Sec.~\ref{sec:dissipon}. We set}
\begin{equation}\label{eq:G1}
\Tilde g_j :=\sqrt{G_j}^*,
\end{equation}
\Kadd{resulting in} $c_j = r_j/\sqrt{G_j}$. This choice has the nice property that with the matrix $P$, now given by $P_{jk} = \frac{r_jr_k^*}{\lambda_j + \lambda_k^*}\frac{1}{\sqrt{G_j}}\frac{1}{\sqrt{G_k}^*}$, we find
\begin{equation}\label{eq:G2}
    (P\bvec g)_j :=\frac{G_j}{\sqrt{G_j}}=\sqrt{G_j}.
\end{equation}
The relevance of this particular choice becomes
clear later when we derive the HEOM equations from the pseudomode approach.

\subsection{Corresponding pseudomode model}
We explained earlier how to transform a general OU process into one that can be represented in terms of a pseudomode model. All we need is the transformation matrix $V$ of \eqref{eq:vandp}, \eqref{eq:diagp}, \eqref{eq:defv}, which here arises from diagonalizing the matrix $P$ with $P_{jk} = \frac{c_j c_k^*}{\lambda_j + \lambda_k^*}$. Then the pseudomode Hamiltonian and the Lindbladians are determined from the matrices $h$ and $\Gamma^\dagger$, which now read, according to Eqs.~\eqref{eq:vcond}, \eqref{eq:gTrafo}
\begin{equation}\label{eq:pseudomodeParametersSI}
    \begin{split}
        h & = \frac{1}{2\i}(V\lambda V^{-1} - \mbox{h.c.})\\
    \Gamma^\dagger & = VB = V \bvec{c}\bvec{\phi}^\dagger = U{\mathcal{D}}^{-1/2}W^\dagger\bvec{c}\bvec{\phi}^\dagger,\\
    \bvec g &= (V^{-1})^\dagger\bvec{\Tilde g},
    \end{split}
\end{equation}
where $\lambda$ is the diagonal matrix with entries $\lambda_j$ and $\Tilde g_j :=\sqrt{G_j}^*$. We set $\bvec{e} = U{\mathcal{D}}^{-1/2}W^\dagger\bvec{c}$ and $\kappa=\bvec{e}^\dagger\bvec{e}$ its norm. Then, due to the unitary freedom $U$, the normalized vector 
$\hat{\bvec{e}} = \bvec{e}/\sqrt{\kappa}$ can point in any desired direction. An interesting choice is $\hat{\bvec{e}}=\bvec{\phi}=\hat{\bvec{e}}_1$, the first basis vector. Then $\Gamma^\dagger = \kappa \hat{\bvec{e}}_1\hat{\bvec{e}}_1^\dagger$, as mentioned in the main text, and only the first pseudomode is damped. 

\subsection{Freedom of choice for the Lindbladian}
The parameters for pseudomode model given in Eq.~\eqref{eq:pseudomodeParametersSI} are not unique, meaning there are generally multiple pseudomode models that result in the same BCF.
We have already highlighted a few of these freedoms along the way, namely the freedom to rescale $\bvec{\tilde g}$ (Eq.~\eqref{eq:G1}) or the unitary freedom of choosing $U$ in Eq.~\eqref{eq:pseudomodeParametersSI}. Both of these freedoms can already be seen from Eq.~\eqref{eq:pseudomodeBCFelegant}, but there is yet another freedom we have not yet discussed.\\

We have shown in Sec.~\ref{ssec:kernelCondition} that it always possible to find a vector $\bvec r$ such that $G_j = \sum_kr_jr_k^*/(\lambda_j+\lambda_k^*)$. However, this vector is not unique and more generally we may also find representations of $G_j$ in terms if a positive matrix $R_{jk}$ if 
\begin{equation}
    \sum_k\frac{R_{jk}}{\lambda_j+\lambda_k^*} = \sum_k\frac{r_jr_k^*}{\lambda_j+\lambda_k^*}.
\end{equation}
Since $R$ is positive, we can find $BB^\dagger = R$ and it is straight forward to verify that the processes 
\begin{equation}
    \dot z_j(t) = -\lambda_j z_j(t) + \sum_{k} B_{jk}/\sqrt{G_j} \xi_k(t),\;\;\;j=1,\ldots,N\, ,
\end{equation}
generate the desired correlation function. 
Crucially $B$ is no longer rank one and thus multiple pseudomodes will be damped.
The freedom of changing $r_kr_j^* \to R_{kj}$ together with the unitary freedom in choosing $V,\,\bvec{\tilde g}$ discussed earlier generates a large set of Lindbladians, which all lead to the same BCF.
Exploring how to find the most numerically efficient Lindbladian from this set of equivalent Lindbladians is a task mostly left for future work.\\

Here, we will restrict ourselves to a statement that uses the parameters \eqref{eq:pseudomodeParametersSI} and is thus restricted to models with a single damped mode. 
Even under this constraint, there is still more freedom in the choice of $U$, which could potentially be utilized to obtain coupling geometries which are suitable numerically. 
A particularly efficient geometry would be a linear chain, which lends itself well to matrix product state techniques \cite{Schollwock2011Jan}.
However, we show in the following that it is generally not possible to find a $U$ which results in the geometry of a chain of $N$ modes with only the last mode being damped.
To obtain the typical chain geometry, the last, damped pseudomode must not couple to the system and thus we find from Eq.~\eqref{eq:pseudomodeParametersSI} the condition
\begin{equation}
    \sum_l \left(V^{-1}\right)^\dagger_{1l}\tilde g_l = \sum_l(U\mathcal{D}^{1/2}W^\dagger)_{1l}\tilde g_l = 0.
\end{equation}
With our choice of $\hat{\bvec e} = \hat{\bvec e}_1$, we find that the first column of $U$ is proportional to $(\mathcal{D}^{-1/2}W^\dagger\bvec c)^*$. Using Eq.~\eqref{eq:gTilde_rRelation} we can reformulate the above condition (for any choice of $\tilde{\bvec g}$) as
\begin{align}
    \sum_l(U\mathcal{D}^{1/2}W^\dagger)_{1l}\tilde g_l =& \bvec c^\dagger W\mathcal{D}^{-1/2}\mathcal{D}^{1/2}W^\dagger\bvec{g},\\
    =& \sum_j r_j^*.
\end{align} 
It is now easy to find physical BCFs, where the additional condition $\sum_kr_k=0$ makes it impossible to find a representation of the form \eqref{eq:exponentialcorrelation2}.  
For example with $N=2$, it is straightforward to verify that the BCF with $\lambda_1 = 1+\i$, $\lambda_2 = 2+\i$, $G_1=1$ and $G_2 =2$ can't be realized by two pseudomodes with $r_1=-r_2$.
Therefore, such a chain-like geometry with one damped mode will in general require more modes than necessary to represent a given BCF.

\section{Dissipon transformation}\label{sec:dissipon}
Now assume that we have constructed the pseudomode model -- as explained above -- corresponding to a physical bath correlation function of the usual exponential form \eqref{eq:exponentialcorrelation2}. Thus, the parameters of that pseudomode model Eq.~\eqref{eq:pseudomodeParametersSI}, including $V,\,\lambda_i$ are known.
In the following we explicitly derive the HEOM and HOPS equations from the dissipon transformation of the pseudomode Hamiltonian 
Eq.~\eqref{eq:SIpseudomodeOU}.
Recall that the Hamiltonian ${H}_{\mathrm{pm,tot}}$ acts on the total Hilbert space containing the system and the environment $\mathcal{H}_{\mathrm{sys}}\otimes\mathcal{H}_{\mathrm{env}}$, where the latter is composed of the pseudomodes and their respective Markovian baths.
The pseudomode Lindblad equation Eqn.(6) arises from an effective vacuum environment initially, such that the total initial state is
\begin{equation}
    |\Psi_0\rangle = |\psi_0\rangle|\mathrm{vac}\rangle_{\mathrm{env}}.
\end{equation}

The ensuing dynamics of the full system-environment state
is determined by the Schrödinger equation
\begin{equation}\label{eq:Psi_cQED}
    \i\partial_t |\Psi(t)\rangle = {H}_{\mathrm{pm,tot}}\left(\bvec{a}(t),\bvec{a}^\dagger(t)\right) |\Psi(t)\rangle\,,
\end{equation} 
where we emphasize the dependence of the Hamiltonian on the time dependent pseudomodes.
Alternatively, we can introduce the unitary propagator $U(t)$ such that
\begin{equation}\label{eq:Psi_cQED2}
    |\Psi(t)\rangle=U(t) |\psi_0\rangle|\mathrm{vac}\rangle_{\mathrm{env}}.
\end{equation}
Ultimately, we are interested in the reduced system state,
\begin{equation}\label{eq:reduced_state}
    \rho_{\mathrm{sys}}(t) = \ptr{\mathrm{env}}{|\Psi(t)\rangle\langle\Psi(t)|},
\end{equation}
that satisfies the pseudomode Lindblad eqn. (6).
We highlight two straightforward observations: first, note that \eqref{eq:SIpseudomodeOU} implies that at equal times, $a_k(t)$ and their time derivatives commute:
\begin{equation}\label{eq:a_adot_commutation}
    [\dot a_k (t), a_j(t)] = 0.
\end{equation}
Therefore, the time derivative of whatever function $f(a_k(t))$ of that operator is given by the usual chain rule $\frac{d}{dt} f(a_k(t)) = f'(a_k(t))\cdot\dot a_k (t)$, or with \eqref{eq:SIpseudomodeOU}
\begin{equation}\label{eq:time_derivative}
    \frac{d}{dt} f(a_k(t)) =  f'(a_k(t)) \left(\sum_{k'}-(ih_{k,k'} + \frac{1}{2}(\Gamma^\dagger\Gamma)_{k,k'})a_{k'} + \Gamma^\dagger_{k,k'}\xi_{k'}(t)\right).
\end{equation}

Second, as the propagator $U(t)$ at time $t$, from integrating its Schr\"odinger equation, can be written in the form $U(t) = \mathbbm{1} - \i\int_0^tds\,H_{\mathrm{pm}}\left(\bvec{a}(s),\bvec{a}^\dagger(s)\right) U(s)$, it is clear that $U(t)$ depends on the white noise process $\bvec{\xi}(s)$ with arguments $s<t$ only. We can thus conclude from (\ref{eq:SIoperatorwhitenoise}) that
\begin{equation}\label{eq:xi_U_t_commutation}
    [\xi_k(t), U(t)] = 0.
\end{equation}
A more rigorous argument would involve operator stochastic calculus with operator Ito-increment $d\xi = \int_t^{t+dt}ds\, \xi(s)$ \cite{Parthasarathy1992}. Both relations, Eqs. \eqref{eq:a_adot_commutation} and \eqref{eq:xi_U_t_commutation} will be useful soon.



To apply the dissipon transformation, we enlarge the Hilbert space further, to 
${\mathcal H}_{\mathrm{sys}}\otimes{\mathcal H}_{\mathrm{env}}\otimes{\mathcal H}_{\mathrm{diss}}$, including the set of $N$ bosonic \emph{dissipon} modes with annihilation (creation) operators $b_k$ ($b_k^\dagger$). In the beam splitter analogy explained in the main text, these represent the "empty", incoming modes of the second input port.
As quantum state in this enlarged Hilbert space we consider $|\Psi(t)\rangle\otimes\mathrm{|vac}\rangle_\mathrm{diss}$, where $\mathrm{|vac}\rangle_\mathrm{diss}$ is the dissipon vacuum (the "empty" port).
We now apply the dissipon transformation 
\begin{align}
    |\Phi(t)\rangle =& D(t)|\Psi(t)\rangle\mathrm{|vac}\rangle_\mathrm{diss},\\
    D(t) =&\exp\left(\sum_{jk}b_j^\dagger V_{jk}^{-1}a_k(t)\right) = \exp\left(\bvec{b}^\dagger V^{-1}\bvec {a}(t)\right),\label{eq:t_dissipon_transformation}
\end{align}
where $V$ corresponds to the Ornstein-Uhlenbeck transformation of the previous sections.
We thus mirror the pseudomodes $\bvec a(t)$ into the static dissipon modes $\bvec b$ in beam splitter style, entangling the pseudomodes with the dissipons. The dynamics of this system-pseudomode-dissipon state $|\Phi(t)\rangle$ will lead us to HEOM and HOPS.
Importantly, as easily seen from \eqref{eq:t_dissipon_transformation}, the original state $|\Psi(t)\rangle$ of the $\bvec a$-modes in (\ref{eq:Psi_cQED}) can be obtained from the entangled state by projection onto the dissipon vacuum,
\begin{equation}\label{eq:root_projection}
    \;_\mathrm{diss}\!\langle\mathrm{vac}|\Phi(t)\rangle = |\Psi(t)\rangle.
\end{equation}
In addition, it is straightforward to confirm the transformation rules
\begin{eqnarray}\label{eq:dissipon_rules}
   \bvec{a}(t) \rightarrow  &  D(t) \bvec{a}(t) D^{-1}(t) &= \bvec{a}(t), \\\nonumber
    \bvec{a}^\dagger(t)  \rightarrow  &  D(t) \bvec{a}^\dagger(t) D^{-1}(t) & = \bvec{a}^\dagger(t) + \bvec{b}^\dagger V^{-1}, \\\nonumber
    \bvec{b} \rightarrow   &  D(t) \bvec{b} D^{-1}(t) & = \bvec{b} - V^{-1}\bvec{a}(t), \\ \nonumber
    \bvec{b}^\dagger \rightarrow  &  D(t) \bvec{b}^\dagger D^{-1}(t) & = \bvec{b}^\dagger,
\end{eqnarray}
which leads to a further interesting observation.
Since $\bvec b$ annihilates the dissipon vacuum, $b_k|\mathrm{vac}\rangle_\mathrm{diss} = 0$, we find from (\ref{eq:dissipon_rules}) $(\bvec b-V^{-1}\bvec a(t))|\Phi\rangle =  D \bvec b D^{-1}|\Phi\rangle = D \bvec b|\Psi\rangle |\mathrm{vac}\rangle_\mathrm{diss} = 0$ or
\begin{equation}\label{eq:ab_annihilation}
 \bvec{a}(t)|\Phi\rangle = V\bvec{b}|\Phi\rangle,
\end{equation}
reflecting the mirroring beam splitter property of the dissipon transformation $D$.\\
Let us now turn to dynamics.
The equations of motion for the dissipon transformed state are given by
\begin{equation}\label{eq:dtphi}
   \partial_t |\Phi_t\rangle = \left(D(t)(\partial_t U(t))+ (\partial_t D(t)) U(t) \right)\,|\psi_0\rangle |\mathrm{vac}\rangle_\mathrm{env}|\mathrm{vac}\rangle_\mathrm{diss}.
\end{equation}
The first term contains the dissipon-transformed Hamiltonian $D(t)(\partial_t U(t))|\psi_0\rangle|\mathrm{vac}\rangle|\mathrm{vac}\rangle_\mathrm{diss} = -\i D(t) H_{\mathrm{pm,tot}} D^{-1}(t) |\Phi_t\rangle$.
Using the relations in Eq.~\eqref{eq:dissipon_rules} a Hamiltonian $H_{\mathrm{pm,tot}}(\bvec a (t),\bvec a^\dagger(t))$ in the Hilbert space of the pseudomodes $\bvec a$ is transformed to the non-Hermitian
\begin{equation}\label{eq:dissipon_H}
    H_{\mathrm{pm,tot}}(\bvec a (t),\bvec a^\dagger(t)) \rightarrow D(t) H_{\mathrm{pm,tot}}(\bvec a(t),\bvec a^\dagger(t)) D^{-1}(t) = H_{\mathrm{pm,tot}}(\bvec a(t),\bvec a^\dagger(t) + \bvec b^\dagger V^{-1}),
\end{equation}
now acting on the extended Hilbert space of both set of modes, $\bvec a (t)$ and $\bvec b$. When acting on the states, we can further use \eqref{eq:ab_annihilation} to write
\begin{equation}\label{eq:dissipon_Hpsi}
    H_{\mathrm{pm,tot}}(\bvec a (t),\bvec a^\dagger(t))|\Psi(t)\rangle \rightarrow  H_{\mathrm{pm,tot}}(\bvec a(t),\bvec a^\dagger(t) + \bvec b^\dagger V^{-1})|\Phi(t)\rangle =  H_{\mathrm{pm,tot}}(V\bvec b,\bvec a^\dagger(t) + \bvec b^\dagger V^{-1})|\Phi(t)\rangle.
\end{equation}
The second term of the time derivative in \eqref{eq:dtphi} simplifies by noting that in $(\partial_t D(t)) U(t)$ the white noise operator $\bvec{\xi}(t)$ from \eqref{eq:time_derivative} acts on $U(t)$. These two operators commute (Eq.~\eqref{eq:xi_U_t_commutation})), and since the white noise then annihilates the environmental vacuum, $\xi_k(t)|\mathrm{vac}\rangle_{\mathrm{env}}=0$, the operator white noise term can be neglected altogether.
In total we find after a slight rearrangement,
\begin{align*}
   \partial_t |\Phi_t\rangle  =& \Bigg(   -\i D(t)H_{\mathrm{pm,tot}}\left(\bvec a(t),\bvec a^\dagger(t)\right)D^{-1}(t)  \notag\\
   &+\sum_{jkl} b_j^\dagger V_{jk}^{-1}\left(-\left(\frac{1}{2}(\Gamma^\dagger\Gamma)_{kl} + \i h_{kl}\right)a_l \right)D(t)\Bigg)U(t)|\psi_0\rangle |\mathrm{vac}\rangle_\mathrm{env}|\mathrm{vac}\rangle_\mathrm{diss},\notag\\ \nonumber
    =& \left(   -\i H_{\mathrm{pm,tot}} \left(V\bvec{b},\bvec{a}^\dagger(t) + \bvec b^\dagger V^{-1}\right) - \bvec{b}^\dagger \lambda \bvec{b}  \right)\, |\Phi_t\rangle.
\end{align*}
Here, we used $\left(\frac{1}{2}\Gamma^\dagger\Gamma +\i h\right) = V\lambda V^{-1}$ from \eqref{eq:vcond}, and again Eq. \eqref{eq:ab_annihilation}. It is remarkable to note that operator ordering issues in $H_{\mathrm{pm,tot}}(\bvec a (t),\bvec a^\dagger(t))$ do not occur under the transformation \eqref{eq:dissipon_Hpsi}, as the replacements satisfy the very same commutator relations:
\begin{equation}
    [(V\bvec{b})_k,(\bvec{a}^\dagger(t) + \bvec b^\dagger V^{-1})_\ell] = \sum_{nm}[V_{kn} b_n, b^\dagger_m V^{-1}_{m\ell}] = \delta_{k\ell} = [a_k(t),a_\ell^\dagger(t)].
\end{equation}
With the actual form of the total pseudomode Hamiltonian from \eqref{eq:SIpseudomodeOU}, we finally arrive at the dissipon-transformed Schr\"odinger equation. We find a non-unitary dynamics in the extended Hilbert space of system, environment, and dissipons, $\cHsys\otimes\cHenv\otimes{\cal{H}}_\mathrm{diss}$, given by
\begin{align}
        \partial_t |\Phi_t\rangle
   & = \Bigg[-\i \left(\Hsys + S\otimes A^\dagger(t) + S\otimes \sum_{kj} (\tilde g_k^* V_{kj}b_j+b_j^\dagger V_{jk}^{-1}\tilde g_k)\right)- \sum_j\lambda_j  b_j^\dagger b_j \Bigg]\, |\Phi_t\rangle,\\ \notag
    & = \Bigg[-\i \left(\Hsys + S\otimes A^\dagger(t) + S\otimes \sum_{j} (g_j^* b_j+(P\bvec{g})_jb_j^\dagger)\right)- \sum_j\lambda_j  b_j^\dagger b_j \Bigg]\, |\Phi_t\rangle,    
\end{align}
From the first to the second line we transformed back to the original parameters --
recall that $\tilde{\bvec g}=(V^{-1})^\dagger\bvec{g}$, and $(V^\dagger V)^{-1} = P$.
Finally, we make use of the specific "HEOM"-choice of the couplings $\bvec{g}$ 
as explained in \eqref{eq:G1}, \eqref{eq:G2} to arrive at an interesting Schr\"odinger-type equation
\begin{align}\label{eq:dissipon_pseudomodeSI}
  \partial_t |\Phi_t\rangle
     & = \Bigg[-\i \left(\Hsys + S\otimes A^\dagger(t) + S\otimes \sum_{j} \sqrt{G_j} \left(b_j+b_j^\dagger\right)\right)- \sum_j\lambda_j  b_j^\dagger b_j \Bigg]\, |\Phi_t\rangle.
\end{align}

Let us comment on this result: the coupling to the $N$ dissipon modes replaces the original coupling to the infinite environment. Those terms are expressed solely using the parameters of the exponential bath correlation function $G_j, \lambda_j$, abandoning all reference to the pseudomode model we constructed for deriving them.
The dissipons themselves are unusual quasiparticles, with energies $\operatorname{Im}(\lambda_j)$ and decay rates $\operatorname{Re}(\lambda_j)$ reflecting the decaying modes with its non-Hermitian "Hamiltonian" $-\i \lambda_jb_j^\dagger b_j$. Moreover, recall that the coupling constants $\sqrt{G_j}$ are complex valued, in general, so the system-dissipon interaction Hamiltonian is non-Hermitian, too. Yet, the original, infinite-dimensional pseudomode environment is still present through another, non-Hermitian term $S\otimes A^\dagger(t)$ that is a remnant of the initial coupling of the system to the infinite, physical pseudomode environment. 
It might seem very confusing why doubling the Hilbert space and considering the more complicated Schr\"odinger equation \eqref{eq:dissipon_pseudomodeSI} should be of any use compared to the original pseudomode equation. 
However, one should not forget that we are ultimately interested in the reduced state \eqref{eq:reduced_state}, which we can obtain by taking the trace over the pseudomode environment (that is both the pseudomodes and their respective Makovian baths) first,
\begin{equation}\label{eq:systemDissiponState}
    \rho_{SD}(t) = \ptr{\mathrm{env}}{|\Phi(t)\rangle\!\langle\Phi(t)|}.
\end{equation}
Thereafter, the system-dissipon state needs to be projected onto the dissipon vacuum $\rho_{\mathrm{sys}} = _\mathrm{diss}\langle\mathrm{vac|}\rho_{SD}\mathrm{|vac}\rangle_\mathrm{|vac}\rangle_\mathrm{diss}$.\\
We show below that, depending on how the trace in Eq.~\eqref{eq:systemDissiponState} is taken, the different hierarchical methods emerge naturally.

\subsection{HEOM}
Taking Eq.~\eqref{eq:systemDissiponState} at face value we can write
\begin{equation}
    \begin{split}
        \dot \rho_{SD}(t) =& -\i[H_{sys},\rho_{SD}] -i S\ptr{env}{A^\dagger(t)\rho_{SD}} +i \ptr{env}{\rho_{SD}A(t)}S- \sum_{j=1}^{N_{exp}} \lambda_jb_j^\dagger b_j\rho_{SD} + \lambda_j^*\rho_{SD} b_j^\dagger b_j\\
        & -\i \sum_{j=1}^{N_{exp}}\sqrt{G_j}S b_j \rho_{SD}- \sqrt{G_j}^{*} \rho_{SD}S b_j^\dagger + \sqrt{G_j}S b_j^\dagger\rho_{SD} - \sqrt{G_j}^{*}\rho_{SD}S b_j.
    \end{split}
\end{equation}
Under the trace we can now replace the pseudomode operators acting from the left (right) with dissipon operators acting from the right (left), because
\begin{equation}
\begin{split}
    \ptr{env}{ |\Phi_t\rangle\langle\Phi_t|a_k(t)} =& \ptr{env}{a_k(t)|\Phi_t\rangle\langle\Phi_t|} = \ptr{env}{\sum_jV_{kj}b_j|\Phi_t\rangle\langle\Phi_t|} = \sum_jV_{kj}b_j\ptr{env}{|\Phi_t\rangle\langle\Phi_t|}\\
    =&  \sum_jV_{kj}b_j \rho_{SD}.
\end{split} 
\end{equation}
From the second to the third term we have made use of the replacement rule in Eq.~\eqref{eq:ab_annihilation}. 
Clearly, an analogous relation holds for the $a_k^\dagger$ operators acting from the left.
We then arrive at the HEOM equation (5) in the main text
\begin{equation}
    \begin{split}
        \partial_t  \rho_{SD}(t) =& -\i \left[\Hsys,\rho_{SD}\right]  - \sum_{j=1}^{N} \left(\lambda_jb_j^\dagger b_j\rho_{SD} + \lambda_j^*\rho_{SD} b_j^\dagger b_j\right) -i \sum_{j=1}^{N} \left(\sqrt{G_j}\left[S, b_j\rho_{SD}\right] + \sqrt{G_j}^{*}\left[S,\rho_{SD} b_j^\dagger\right]\right)\\
        & -\i \sum_{j=1}^{N}\left(\sqrt{G_j}Sb_j^\dagger\rho_{SD}  - \sqrt{G_j}^{*} \rho_{SD} b_jS\right).
    \end{split}
\end{equation}

\subsection{Linear HOPS}
Alternatively, we may take the environmental trace in Eq.~\eqref{eq:systemDissiponState} explicitly in a basis of (Bargmann) coherent states $\|\bvec{z}\rangle = \exp\left(\sum_{j=1}^{N}z_ja_j^\dagger\right)\exp\left(\sum_{k,\lambda}^{N}z_{k,\lambda}d_{k,\lambda}^\dagger\right)\mathrm{|vac}\rangle_{env}$, where the vector $\bvec z$ now contains the labels for all pseudomodes $z_j$ as well as their environments $z_{k,\lambda}$. 
The trace can then be written as
\begin{equation}\label{eq:bargmannTrace}
    \rho_{SD}(t) = \int \left(\prod_{k,\lambda}\frac{d^2z_{k,\lambda}}{\pi}\right)\left(\prod_{j}\frac{d^2z_j}{\pi}\right) e^{-|\bvec z|^2}\,\langle \bvec z \|\Phi(t)\rangle\!\langle\Phi(t)\|\bvec z\rangle,
\end{equation}
which provides a stochastic unravelling of the system-dissipon-state as a Gaussian mixture of pure states
\begin{equation}\label{eq:sys_dissipon_HOPS_state}
    |\phi_t(\bvec z ^*)\rangle := \langle \bvec z \|\Phi(t)\rangle.
\end{equation}
Considering the integral in (\ref{eq:bargmannTrace}) as an average over Gaussian random numbers $z_j,\ z_{k,\lambda}$ in $\bvec z$, the system-dissipon state is an expectation value $\mathbb{E}[\ldots]$ of stochastic pure states,
\begin{equation}\label{eq:sys_dissipon_HOPS2}
    \rho_{SD}(t) = \mathbb{E}[ |\phi_t(\bvec z ^*)\rangle\langle \phi_t(\bvec z ^*)| ].
\end{equation}
Their time evolution is easily determined from the fundamental dissipon equation \eqref{eq:dissipon_pseudomodeSI} by replacing the remaining environmental Ornstein-Uhlenbeck operator $A^\dagger(t)$  by the corresponding $c$-number expression through $\langle \bvec z \|(-\i A^\dagger(t)) =: z_t^* \langle \bvec z \|$, such that the closed system-dissipon evolution equation reads
\begin{align}\label{eq:dissipon_LHOPS}
   \partial_t |\phi_t(z^*)\rangle
   & = \left[  -\i \Hsys + Sz_t^*  -\i\sum_{j}S (\sqrt{G_j}b_j+\sqrt{G_j}b_j^\dagger )- \sum_j\lambda_j  b_j^\dagger b_j\right]|\phi_t(z^*)\rangle.
\end{align}
\Kadd{As discussed in Sec.~\ref{sec:OUprocesses},} the c-number time-dependent function $z_t^*$ inherits the correlation of the underlying Ornstein-Uhlenbeck processes, i.e. it is a c-number Gaussian stochastic process with
\begin{equation}\label{eq:ou_correlations_z}
\mathbb{E} [z_t  z_s^*]
= \sum_k G_ke^{-\lambda_k(t-s)}\;\;(t\ge s\ge 0).
\end{equation}
Upon expanding the dissipon sector in number states,
\begin{equation}\label{eq:HOPS_expansion}
   |\phi_t(z^*)\rangle = \sum_{n=0}^\infty |\psi_t^{(n)}(z^*)\rangle |n\rangle_\mathrm{diss},
\end{equation}
the hierarchy of system states $|\psi_t^{(n)}(z^*)\rangle$ satisfies the usual (linear) HOPS. From Eq.~\eqref{eq:sys_dissipon_HOPS2} it is clear that the physical reduced state of the system is obtained from the average over the zeroth order hierarchy state, or the projection onto the dissipon vaccum
\begin{equation}
    \rho_{sys} = \langle\mathrm{vac|}\rho_{SD}\mathrm{|vac}\rangle = \mathbb{E}[\langle\mathrm{vac|}\phi_t(\bvec z^*)\rangle\!\langle\phi_t(\bvec z^*)\mathrm{|vac}\rangle].
\end{equation}
Finally, we note that in the original HOPS \cite{HOPS, HOPS_Richard} the stochastic process has the exact correlation function $\alpha(t-s)$, whereas here we find the exponential correlation function $\alpha_{exp}(t-s) \approx \alpha(t-s)$. 
Strictly speaking, this should make the original HOPS more accurate than Eq.~\eqref{eq:dissipon_LHOPS} derived from the pseudomode approach. However, given the level of error excepted by approximating $\alpha(t-s)\approx \alpha_{exp}(t-s)$ (which is also done in the original HOPS) this is unlikely to be significant.

\subsection{Non-linear HOPS}
The linear HOPS equation (\ref{eq:dissipon_HOPS}) can be transformed into its far superior non-linear version by the usual time-dependent change of environmental coherent state labels \cite{NMQSD}. 
The non-linear, time-dependent transformation to the non-linear HOPS arises from considering the Husimi-Q-function of the pseudomode environment, 
$Q_t(\bvec z,\bvec z^*) = \frac{e^{-|\bvec z|^2}}{\pi} \langle \bvec z \|\ptr{sys}{|\Psi(t)\rangle\!\langle\Psi(t)|}\|\bvec z\rangle$, with $|\Psi(t)\rangle$ from Eq.~\eqref{eq:Psi_cQED} 
As explained elsewhere \cite{richardThesis,Boettcher2024Mar}, the evolution equation for $Q_t(\bvec z,\bvec z^*)$ turns into a Liouville-type flow equation. In our case, based on the pseudomode Hamiltonian \eqref{eq:fundamentalHCQED}, the corresponding evolution equations for the labels read
\begin{align}\label{eq:cQED_flow}
    \dot z_j^* & = \i \langle S \rangle_t\sum_kg_k\left(e^{-(ih + 1/2\Gamma^\dagger\Gamma) t}\right)_{kj} \\ \nonumber
    \dot z_{j,\lambda}^* & = \i\langle S \rangle_t\sum_{k,k'} g_k\int_0^t\dd s \left(e^{-(ih + 1/2\Gamma^\dagger\Gamma) (t-s)}\right)_{kk'}\Gamma_{k'j}\eta_{j\lambda} e^{-\i\omega_{j,\lambda}s}.
\end{align}
Here, the expectation value 
\begin{equation}
    \langle S\rangle_t = \frac{\langle\phi_t(\bvec z^*)|\left( S \otimes |\mathrm{vac}\rangle_\mathrm{diss} \;_\mathrm{diss}\!\langle\mathrm{vac}|\right)|\phi_t(\bvec z^*)\rangle}{\langle\phi_t(\bvec z^*)|\mathrm{vac}\rangle_\mathrm{diss} \;_\mathrm{diss}\!\langle\mathrm{vac}|\phi_t(\bvec z^*)\rangle} = \frac{\langle\psi_t^{(0)}(\bvec z^*)|S|\psi_t^{(0)}(\bvec z^*)\rangle}{\langle\psi_t^{(0)}(\bvec z^*)|\psi_t^{(0)}(\bvec z^*)\rangle}
\end{equation}
is the usual, normalized quantum expectation value of the system coupling agent with respect to the physical ($n=0$) dissipon-vacuum state $ |\psi_t^{(0)}(\bvec z^*)\rangle$ in (\ref{eq:HOPS_expansion}).

Upon integrating Eqs.~\eqref{eq:cQED_flow} we find the usual replacement rule for the stochastic process,
\begin{equation}\label{eq:girsanov}
    z_t^* \rightarrow \tilde{z}_t^* = z_t^* + \int_0^t ds\, \alpha_{exp}^*(t-s) \langle S\rangle_s,
\end{equation}
exhibiting the exponentially decaying Ornstein-Uhlenbeck correlation function under the memory integral. Moreover, once the coherent state labels become time dependent, we need a comoving ("convective") time derivative i.e.
\begin{equation}
    \frac{d}{dt} |\phi_t(\bvec z^*(t))\rangle = \partial_t |\phi_t(\bvec z^*(t))\rangle + \sum_j\partial_{z_j^*}|\phi_t(\bvec z^*(t))\rangle\dot z_j^* + \sum_{k,\lambda} \partial_{z_{k,\lambda}^*}|\phi_t(\bvec z^*(t))\rangle\dot z_{k,\lambda}^*.
\end{equation}
As $\partial_{z_j^*}|\phi_t(z^*)\rangle = \partial_{z_j^*}\langle \bvec z\|\Phi_t\rangle = \langle \bvec z\|a_j|\Phi(t)\rangle$, and similarly for $\partial_{z_{k,\lambda}^*}|\phi_t(\bvec z^*)\rangle = \langle \bvec z\|d_{k,\lambda}|\Phi(t)\rangle$, we find using (\ref{eq:cQED_flow}) the simple relation
\begin{equation}\label{eq:convection}
    \partial_{z_k^*}|\phi_t(\bvec z^*(t))\rangle\dot z_k^* + \sum_\lambda \partial_{z_{k,\lambda}^*}|\phi_t(\bvec z^*(t))\rangle\dot z_{k,\lambda}^* = \i g_k\langle S\rangle_t\langle \bvec z\|a_k(t)|\Phi(t)\rangle,
\end{equation}
with the pseudomodes $a_k(t)$ from Eq.~\eqref{eq:SIpseudomodeOU}. Yet $a_k(t)|\Phi(t)\rangle = \sum_j V_{kj}b_j|\Phi(t)\rangle$ according to the dissipon transformation rules and thus, combining the latest findings, we can express the comoving time derivative for the system-dissipon state $ |\phi_t\rangle := |\phi_t(z^*(t))\rangle$ in comoving coherent state labels with the help of the dissipon annihilation operator as
\begin{equation}
     \frac{d}{dt} |\phi_t\rangle  = \partial_t |\phi_t\rangle + \i\sqrt{G_j}\langle S\rangle_t b |\phi_t\rangle.
\end{equation}

Finally, we combine with the evolution equation of linear HOPS (\ref{eq:dissipon_LHOPS}) and find the non-linear version of the HOPS in the dissipon picture,
\begin{align}\label{eq:dissipon_HOPS}
    \frac{d}{dt} |\phi_t\rangle
   & = \left[  -\i \Hsys + S{\tilde z}_t^*  -\sum_j\i \sqrt{G_j}(S-\langle S \rangle_t) b_j -\i \sqrt{G_j}S\otimes b_j^\dagger  - \lambda_j b_j^\dagger b_j \right]\, |\phi_t\rangle,
\end{align}
where ${\tilde z}_t^*$ is the shifted process from (\ref{eq:girsanov}).

\subsection{nuHOPS}
Clearly, the "effective Hamiltonian" on the right-hand-side of the non-linear HOPS equation (\ref{eq:dissipon_HOPS}) is far from Hermitian, and furthermore a direct Fock-state expansion might not lead to the most efficient numerical representation, especially for highly excited environments. With a further transformation -- very akin to the original dissipon-transformation but much more elementary -- we can achieve a
"near-unitary" HOPS, the {\it nuHOPS} \cite{nuHOPS}: we introduce a new system-dissipon state $|\psi_t\rangle$ through 
\begin{equation}\label{eq:t_dissipon_transformation_nu}
    |\psi_t\rangle =  \exp\{-\sum_k\nu_k(t) b_k^\dagger\}|\phi_t\rangle.
\end{equation}
This transformation amounts to a shift of the dissipon by a time dependent amplitude:
\begin{equation} \label{eq:nuHOPS_shift}
    e^{-\sum_k\nu_k(t) b_k^\dagger}\,b_j\,  e^{\sum_k\nu_k(t) b^\dagger} = b_j+\nu_j(t),
\end{equation}
For the reasons discussed in Ref.~\cite{nuHOPS}, we choose $\nu(t)$ to satisfy the "mean field" equation
\begin{equation}\label{eq:mean_field}
    \dot\nu_k = -\lambda_k\nu_k -\i\langle S\rangle_t,
\end{equation}
with initial condition $\nu_k(0)=0$. Other choices are possible, yet with this $\bvec \nu(t)$ we have
\begin{equation}
    \nu_k(t) = -\i \int_0^t G_ke^{-\lambda_k(t-s)}\langle S\rangle_s,
\end{equation}
an expression which readily recombines to the shifted process in Eq.~\eqref{eq:girsanov}.\Kadd{It can now} be written in terms of $\bvec \nu(t)$ as:
\begin{equation}\label{eq:girsanov2}
    z_t^* \rightarrow \tilde{z}_t^* = z_t^* -\i\sum_k \nu_k^*(t).
\end{equation}

Note that the relevant dissipon-vacuum-projected state is not affected by this transformation,
\begin{equation}
   |\psi_t^{(0)}\rangle := _\mathrm{diss}\!\langle\mathrm{vac}|\psi_t\rangle =   _\mathrm{diss}\!\langle\mathrm{vac}|\phi_t\rangle, 
\end{equation}
so that the required reduced system state can be obtained from these transformed states as the vacuum-projected, normalized ensemble mean
\begin{equation}\label{eq:nuHOPS_ensemble}
    \rho_t =   \mathbb{E}\left[ \frac{\;|\psi_t^{(0)}\rangle\langle \psi_t^{(0)}|}{\langle \psi_t^{(0)}|\psi_t^{(0)}\rangle}\right],
\end{equation}
as requested.

The evolution equation of the transformed states $|\psi_t\rangle$ is easily found: taking the time derivative in (\ref{eq:t_dissipon_transformation_nu}), and considering the shift property (\ref{eq:nuHOPS_shift}),
together with the evolution equation (\ref{eq:dissipon_HOPS}) for $|\phi_t\rangle$ we find the {\it near-unitary} nuHOPS equation
\begin{align}\label{eq:dissipon_nuHOPS}
    \frac{d}{dt} |\psi_t\rangle
   & = -\i \left[  \Hsys   + S\sum_k(\nu_k(t)+\nu_k^*(t))+ \sum_k(S-\langle S \rangle_t) \otimes (\Tilde G_k^{(1)}b_k+ G_k^{(2)}b_k^\dagger)  +\operatorname{Im}(\lambda_k) b_k^\dagger b_k \right]\, |\psi_t\rangle\\ \nonumber
   & \;\;\;\;+ \left((S-\langle S \rangle_t) z_t^* - \sum_k\operatorname{Re}(\lambda_k)  (b_k^\dagger b_k-\langle b_k^\dagger b_k\rangle)\right)|\psi_t\rangle,
\end{align}
which has to be solved along with the "mean-field" equation (\ref{eq:mean_field}) for the shift $\bvec \nu(t)$. Note that we have written the nuHOPS equation in a norm-preserving way: $\langle\psi_t|\psi_t\rangle = 1$ for all times.

It is the second line only that leads to a non-unitary evolution of $|\psi_t\rangle$, the process $z_t^*$ appearing there is the original Ornstein-Uhlenbeck process, since the shift in (\ref{eq:girsanov}) can been combined with other terms arising from the transformation (\ref{eq:t_dissipon_transformation_nu}) to an overall unitary contribution. 
Most remarkably, the effect of that $\bvec \nu(t)$-transformation (\ref{eq:t_dissipon_transformation_nu}) is not only the taming of non-unitary terms in the usual HOPS: the shift $\nu(t)$ actually ensures that the dissipon will not get highly excited during time evolution, $\langle b \rangle_t \approx 0 $ for all times, which means that only a few dissipon number states (aka hierarchy terms) need to be taken into account. 

Two final remarks: first, in the derivation of (\ref{eq:dissipon_nuHOPS}) we neglected an additional, purely time dependent term which, however, is entirely irrelevant for the determination of the reduced state as it cancels in the normalized projector combination (\ref{eq:nuHOPS_ensemble}) of the $|\psi_t\rangle$. So, strictly speaking, the two states $|\psi_t\rangle$ in (\ref{eq:t_dissipon_transformation_nu}) and (\ref{eq:dissipon_nuHOPS}) differ by an irrelevant time-dependent factor.

Second, and more importantly: one may be tempted to consider the nuHOPS shift-operation (\ref{eq:t_dissipon_transformation_nu}) directly from the start, by changing the original dissipon transformation (\ref{eq:t_dissipon_transformation}) to
\begin{equation}
    D(t) = \exp\left(\sum_{jk}(V_{jk}^{-1}a_k(t)-\mu_j(t)) b_j^\dagger\right),
\end{equation}
as suggested by (\ref{eq:t_dissipon_transformation_nu}).
Then however, the "mean field" equation for $\bvec \mu(t)$ is determined by the ensemble-averaged state $\rho_t$. Here, in nuHOPS, the "mean-field" $\nu(t)$ is obtained for each stochastic pure state $|\psi_t\rangle$ independently, and thus leads to a shift that is tailored and optimized for each trajectory \cite{nuHOPS}.

\end{document}